\documentclass[a4paper,11pt]{article}
\pdfoutput=1
\usepackage{jheppub} 
\usepackage{epsfig}
\usepackage[caption=false]{subfig}
\usepackage{amsmath}
\usepackage{amsfonts}
\usepackage{amssymb}
\usepackage{float}
\usepackage{color}
\usepackage{graphicx}
\usepackage{hyperref}
\usepackage{adjustbox,array}
\hypersetup{}
\usepackage[utf8x]{inputenc} 
\usepackage{epstopdf}
\usepackage{multirow}
\usepackage{slashed}
\usepackage{booktabs}
\usepackage{gensymb}
\usepackage{comment}
\usepackage{diagbox}

\hypersetup{
  colorlinks=true,
  linkcolor=red,
  urlcolor=blue,
  citecolor=red, %gray
  pdfauthor={Name}
}

\newcommand{\alphas}{\alpha_{\text{s}}}

\newcommand{\mlq}{{m_{\text{LQ}}}}
\newcommand{\pwb}{{\textsc{POWHEG-BOX}}}

\newcommand{\mgamc}{{\textsc{MadGraph5\_aMC@NLO}}}

\def\sss{\scriptscriptstyle}
\newcommand{\be}{\begin{equation}}
\newcommand{\ee}{\end{equation}}
\def\bsp#1\esp{\begin{split}#1\end{split}}
\def\bpm{\begin{pmatrix}}
\def\epm{\end{pmatrix}}

\def\lag{{\cal L}}
\def\dR{d_{\sss R}}

\def\eR{\ell_{\sss R}}
\def\eRbar{{\bar \ell}_{\sss R}}
\def\Ll{L_{\sss L}}

\def\QL{Q_{\sss L}}
\def\QLbar{\bar Q_{\sss L}}
\def\uR{u_{\sss R}}
\def\uRbar{{\bar u}_{\sss R}}
\def\uL{u_{\sss L}}
\def\uLbar{{\bar u}_{\sss L}}
\def\dL{d_{\sss L}}
\def\dLbar{{\bar d}_{\sss L}}
\def\nuL{\nu_{\sss L}}

\def\eL{\ell_{\sss L}}

\def\eg{{\it e.g.}}
\def\ie{{\it i.e.}}

\makeatletter
\newcommand\footnoteref[1]{\protected@xdef\@thefnmark{\ref{#1}}\@footnotemark}
\makeatother

\preprint{KIAS-Q22010,  CTPU-PTC-22-24, \begin{flushright} \vspace{-.35cm} CERN-TH-2022-106, KA-TP-17-2022, \end{flushright} \begin{flushright} \vspace{-.45cm} MS-TP-22-18
\end{flushright}
}

\title{Scalar leptoquarks at the LHC and flavour anomalies: a comparison of pair-production modes at NLO-QCD}

\author[a]{Christoph Borschensky}
\author[b]{\!\!, Benjamin Fuks}
\author[c,d]{\!\!, Adil Jueid}
\author[e,f]{\! and Anna Kulesza}

\affiliation[a]{Institut f\"ur Theoretische Physik, Karlsruhe Institute for Technology, 76128 Karlsruhe, Germany}
\affiliation[b]{Laboratoire de Physique Th\'eorique et Hautes \'Energies (LPTHE), UMR 7589, Sorbonne Universit\'e et CNRS, 4 place Jussieu, 75252 Paris Cedex 05, France}
\affiliation[c]{Center for Theoretical Physics of the Universe, Institute for Basic Science (IBS), Daejeon, 34126, Republic of Korea}
\affiliation[d]{Quantum Universe Center, Korea Institute for Advanced Study, Seoul 02455, Republic of Korea}
\affiliation[e]{Institut f\"ur Theoretische Physik, WWU M\"unster, D-48149 M\"unster, Germany}
\affiliation[f]{Theoretical Physics Department, CERN, 1211 Geneva 23, Switzerland}

\emailAdd{christoph.borschensky@kit.edu}
\emailAdd{fuks@lpthe.jussieu.fr}
\emailAdd{adiljueid@ibs.re.kr}
\emailAdd{anna.kulesza@uni-muenster.de}

\abstract{We analyse scalar leptoquark pair production at the LHC with predictions including $t$-channel lepton exchange contributions up to next-to-leading order (NLO) in QCD. In particular, we calculate NLO-QCD predictions for off-diagonal production channels, {\it i.e.}\ channels that involve two different leptoquark eigenstates and are driven solely by diagrams involving Standard Model leptons in the $t$-channel at leading order, as opposed to diagonal channels where a pair of the same leptoquark eigenstate is produced. We find that reliable theoretical predictions for both channels require NLO accuracy. The relative importance of the off-diagonal modes depends strongly on the considered scenario. In a generic model involving $R_2$ and $S_3$ leptoquarks, at large values of the Yukawa couplings off-diagonal contributions initiated by valence quarks can be up to an order of magnitude higher than the diagonal production. However, we also find that in phenomenologically viable scenarios addressing the flavour anomalies off-diagonal production is generally negligible, with a few exceptions of 10\%--30\% of the total rate depending on the treatment of the charm density in the proton.}

\begin{document} 
\maketitle
\flushbottom 

%%%%%%%%%%%%%%%%%%%%%%%%%%%%%%%%%%%%%%%%%%%%%%
\section{Introduction}
\label{sec:intro}
%%%%%%%%%%%%%%%%%%%%%%%%%%%%%%%%%%%%%%%%%%%%%%%

Scalar leptoquarks are scalar bosons beyond the Standard Model which carry both lepton and baryon numbers, and therefore couple simultaneously to leptons and quarks via a Yukawa-type interaction. Originally proposed in models of Grand Unification \cite{Pati:1973uk,Pati:1974yy,Georgi:1974sy,Fritzsch:1974nn,Senjanovic:1982ex,Frampton:1989fu,Murayama:1991ah}, they appear in many extensions to the Standard Model, \eg~in superstring models \cite{Hewett:1988xc}, $R$-parity violating supersymmetric scenarios \cite{Farrar:1978xj,Barbier:2004ez}, composite models \cite{Dimopoulos:1979es,Eichten:1979ah,Farhi:1980xs,Schrempp:1984nj,Lane:1991qh}, neutrino mass models~\cite{Dorsner:2005fq,Babu:2010vp,Babu:2019mfe,Bigaran:2019bqv,Julio:2022ton,Julio:2022bue}, or as mediators in simplified models of dark matter \cite{Baker:2015qna,Choi:2018stw,Guadagnoli:2020tlx,Arcadi:2021cwg,Belanger:2021smw,Baker:2021llj}. Scalar leptoquarks may also offer an explanation \cite{Hiller:2014yaa,ColuccioLeskow:2016dox,Crivellin:2017zlb,Hiller:2018wbv,Crivellin:2020tsz,Angelescu:2021lln,Nomura:2021oeu,Marzocca:2021azj,FileviezPerez:2021lkq,Murgui:2021bdy,Singirala:2021gok,Crivellin:2022mff} to the anomalies appearing in $B$-physics observables~\cite{BaBar:2012obs,Belle:2019oag,BELLE:2019xld,Belle:2019rba,LHCb:2017avl,LHCb:2017smo,LHCb:2019hip,LHCb:2021trn}, as well as to the discrepancy between theoretical predictions \cite{Aoyama:2020ynm} and experimental measurements \cite{Muong-2:2006rrc,Muong-2:2021ojo} related to the anomalous magnetic moment of the muon. In this context, most interesting scenarios feature large Yukawa couplings between leptons, quarks, and leptoquarks.

The ATLAS and CMS collaborations at the Large Hadron Collider (LHC) have so far measured no signals hinting at the existence of scalar leptoquarks, thus pushing the mass limits up to around 1.0--1.8\,TeV~\cite{ATLAS:2019qpq,ATLAS:2020dsf,ATLAS:2020dsk,ATLAS:2021oiz,ATLAS:2021yij,ATLAS:2021jyv,ATLAS:2022ors,ATLAS:2022fho,ATLAS:2022wgt,CMS:2018txo,CMS:2018lab,CMS:2018iye,CMS:2018ncu,CMS:2020wzx,CMS:2021far,CMS:2022nty}. The precise value of the bounds depend on the considered leptoquark model (in particular on the texture in the flavour space of the leptoquark Yukawa couplings), as well as on the production mechanism targeted by the different searches. The most stringent limits originate from LHC direct searches for the pair production of scalar leptoquarks, assuming that the Yukawa couplings are small so that the production is driven purely by QCD interactions. Leptoquark masses are hence constrained to be larger than 1.8~TeV~\cite{ATLAS:2020dsk,CMS:2018ncu,CMS:2021far}, 1.7~TeV~\cite{ATLAS:2020dsk,CMS:2018lab} and 1.4~TeV~\cite{ATLAS:2019qpq,ATLAS:2020dsf,ATLAS:2021oiz,ATLAS:2021yij,ATLAS:2021jyv,CMS:2018iye,CMS:2020wzx} for leptoquarks coupling to first-generation, second-generation and third-generation fermions respectively. Moreover, mixed leptoquark decay patterns into leptons and quarks of different generations have been studied, which has led to mass limits of about 1.7--2.0~TeV for both the electron and muon channels~\cite{ATLAS:2020dsk,ATLAS:2022ors,ATLAS:2022wgt}. The assumption of a small Yukawa coupling is however clearly at odds with possible explanations of lepton flavour anomalies. Consequently, Yukawa coupling-induced contributions ought to be included in studies of leptoquark production at the LHC. Such a large Yukawa coupling assumption also renders other leptoquark production modes relevant, and bounds of 1.4~TeV~\cite{ATLAS:2022fho,CMS:2018txo,CMS:2020wzx,CMS:2022nty} have correspondingly been obtained for the third-generation leptoquark case. These mass limits can nevertheless be weakened or even evaded by reducing the size of the leptoquark Yukawa couplings.

From the theoretical side, calculations for cross sections of scalar leptoquark pair production are available at the next-to-leading order (NLO) in the strong coupling constant $\alphas$ already for some time \cite{Kramer:1997hh,Kramer:2004df}. More recently these fixed-order calculations have been matched with parton showers~\cite{Dorsner:2018ynv,Mandal:2015lca}, and with soft-gluon resummation up to next-to-next-to-leading logarithmic accuracy \cite{Borschensky:2020hot, Borschensky:2021hbo,Borschensky:2021jyk}. In the latter works, in addition to the pure-QCD contribution, also so-called $t$-channel contributions (\ie~diagrams with a $t$-channel lepton exchange that are proportional to the leptoquark Yukawa couplings) have been taken into account. It has been shown that for phenomenologically relevant parameter points, these additional terms can be sizeable and can interplay in an intricate way with soft-gluon corrections. Furthermore, the expected production rates turns out to depend strongly on the chosen leptoquark model and benchmark scenario, as well as on the set of parton distribution functions (PDF) used for the calculations. All these effects need thus to be considered for the best exploitation of scalar leptoquark searches.

Besides pair production of scalar leptoquarks, other relevant processes for leptoquark searches at hadron colliders include their single production in association with a lepton~\cite{Hiller:2018wbv,Alves:2002tj,Dorsner:2014axa,Hammett:2015sea,Mandal:2015vfa} (for which predictions are known at NLO in QCD~\cite{Dorsner:2018ynv}), Drell-Yan di-lepton production with leptoquarks appearing as virtual particles~\cite{Faroughy:2016osc,Raj:2016aky,Greljo:2017vvb,Bansal:2018eha,Schmaltz:2018nls,Fuentes-Martin:2020lea,Haisch:2022lkt,Haisch:2022afh}, and resonant leptoquark production through quark-lepton initial states~\cite{Ohnemus:1994xf,Eboli:1997fb,Buonocore:2020erb,Greljo:2020tgv,Buonocore:2022msy}. Recently, a novel channel related to the pair production of scalar leptoquarks has been introduced~\cite{Dorsner:2021chv}. In certain scenarios with at least two leptoquark eigenstates, there exists the possibility to produce a `diagonal' pair of leptoquarks of the same eigenstate, or to `off-diagonally' produce two different leptoquark states. In contrast to the diagonal production, the off-diagonal channel is, at tree level, purely driven by the strength of the leptoquark Yukawa couplings and independent of the strong coupling constant. In addition, the quark-quark channel opens up, with quarks of possibly unequal flavours. This is particularly suited for LHC studies as one can expect large cross sections if two valence quarks are involved. For instance, it has been shown at leading order (LO) in~\cite{Dorsner:2021chv} that for a one-flavour new physics realisation based on a model including both $S_1$ and $R_2$ leptoquarks, the {\it total} pair production rates  can be enhanced strongly relative to the expectation from pure QCD production, and this even for moderate leptoquark Yukawa couplings. In addition, the off-diagonal production channels have the potential to outweigh the importance of other Yukawa-coupling-dependent search channels, especially for moderate to large Yukawa couplings.

In the present work, we investigate the relevance of the off-diagonal production mechanism in more detail, in particular in the context of models relevant for explaining flavour anomalies. 
Given the impact of higher-order QCD effects on the predictions for scalar leptoquark pair-production reported in~\cite{Borschensky:2020hot, Borschensky:2021hbo,Borschensky:2021jyk}, and generally their relevance for the production of heavy coloured particles in hadronic channels, we increase the precision of the off-diagonal production mechanism of scalar leptoquarks by calculating the associated NLO-QCD corrections. We also compare the off-diagonal production rates with diagonal production through both QCD and  Yukawa coupling-induced $t$-channel mechanisms. The computations are included in our framework of scalar leptoquark pair production for various leptoquark models \cite{Borschensky:2020hot,Borschensky:2021hbo,Borschensky:2021jyk} with the \mgamc{}~\cite{Alwall:2014hca} and \pwb~\cite{Nason:2004rx,Frixione:2007vw,Alioli:2010xd} software. In addition, we comprehensively study the impact of the Yukawa-coupling-driven contributions on inclusive rates for phenomenologically viable leptoquark models that offer a solution to the flavour anomalies.

The structure of this paper is as follows. In section~\ref{sec:models}, we discuss our theoretical setup and the specific benchmark leptoquark scenarios that we study, followed by a brief review of leptoquark pair production at fixed order. We then introduce some computational details in section~\ref{sec:results}, and next analyse  the numerical impact of the off-diagonal production modes on total rates. In this context, we consider a general scenario including $R_2$ and $S_3$ leptoquarks coupling to quarks of the first and second generation. An application of our calculations to phenomenologically viable benchmark scenarios is subsequently presented in section~\ref{sec:BSs}, where we also highlight the role of the charm PDF on production rates. At last, we conclude in section~\ref{ref:conclusions}.

%%%%%%%%%%%%%%%%%%%%%%%%%%%
\section{Theoretical setup}
\label{sec:models}
%%%%%%%%%%%%%%%%%%%%%%%%%%%

%%%%%%%%%%%%%%%%%%%%%%%%%%%%%
\subsection{Phenomenological leptoquark models}
%%%%%%%%%%%%%%%%%%%%%%%%%%%%%

We consider various simplified leptoquark models where the Standard Model (SM) is supplemented by the following species of scalar leptoquarks: $S_1$, $R_2$, and $S_3$. We furthermore assume that these leptoquarks couple only to the SM fermions and not to exotic beyond-the-SM ones. Following the notation of references \cite{Buchmuller:1986zs,Dorsner:2016wpm}, these leptoquarks transform as $S_1:~({\bf 3},{\bf 1})_{\sss -1/3},~R_2:~({\bf 3},{\bf 2})_{\sss 7/6},~S_3:~({\bf 3},{\bf 3})_{\sss-1/3}$ under the $SU(3)_c \otimes SU(2)_L \otimes U(1)_Y$ gauge group. 

Following these transformation rules, the most general gauge invariant and renormalisable Lagrangian involving the $S_1$, $R_2$, and $S_3$ species is given by\footnote{We do not study the scalar potential involving Higgs multiplets and scalar leptoquarks (that can be found, \eg,  in  \cite{Crivellin:2021ejk}), since we are studying scalar leptoquark pair production in simplified models.}
\be\bsp
   \lag_{\rm LQ} = &\ \lag_{\rm kin.} + \lag_{\rm mass} + 
   {\bf y_{\sss 1}^{\sss RR}} \uRbar^c \eR S_1^\dagger
  +{\bf y_{\sss 1}^{\sss LL}} \big(\QLbar^c \!\cdot\! \Ll\big) S_1^\dagger
  +{\bf y_{\sss 2}^{\sss LR}} \eRbar \QL R_2^\dag
 \\ &\ +
   {\bf y_{\sss 2}^{\sss RL}} \uRbar\big(\Ll\!\cdot\!R_2\big)
  +{\bf y_{\sss 3}^{\sss LL}} \big(\QLbar^c\!\cdot\!\sigma_k \Ll\big) \big(S_3^k\big)^\dag
  + {\rm H.c.},
\esp\label{eq:lag}\ee
where we have represented the $S_3$ multiplet as a vector carrying an adjoint $SU(2)_L$ index, namely $S_3^k$ with $k=1,2,3$. An alternative way to write the above Lagrangian could make use of a matrix representation for $SU(2)_L$ triplets. In this case, $S_3 \equiv (S_3)^\alpha{}_{\dot{\alpha}} = \frac{1}{\sqrt{2}} (\sigma_k)^\alpha{}_{\dot{\alpha}} S_3^k$, where the $S_3$ matrix carries a fundamental ($\alpha=1,2$) and an antifundamental ($\dot{\alpha}=1,2$) index of $SU(2)_L$, and where $\sigma_k$ are the usual Pauli matrices. In addition, $\big(\QLbar^c \cdot \Ll\big)$, $\big(\Ll \cdot R_2\big)$, $\big(\QLbar^c \cdot \sigma_k \Ll\big)$ refer to standard $SU(2)_L$ invariant products. In the above Lagrangian, we have suppressed both generation as well as colour indices. All the couplings ${\bf y}$ are $3\times 3$ matrices in the flavour space and which can be complex as well. The first index of any element $y_{ij}$ of a coupling matrix ${\bf y}$ refers to a quark generation while the second index refers to a lepton generation. The quark and lepton multiplets are denoted by  $\QL$ and $\Ll$ (quark and lepton weak isodoublets), $\uR$, $\dR$ and $\eR$ (quark and lepton isosinglets). Finally, all the kinetic and mass terms involving scalar leptoquarks are encoded in $\lag_{\rm kin.}$ and $\lag_{\rm mass}$, respectively. 

The scalar leptoquarks are represented as follows,
\begin{eqnarray}
S_1  = S_1^{(-1/3)} \ , \qquad R_2        = \bpm R_2^{(+5/3)} \\ R_2^{(+2/3)}\epm\ ,\qquad S_3 = \bpm \frac{1}{\sqrt{2}}S_3^{(-1/3)} & S_3^{(+2/3)}\\
              S_3^{(-4/3)} & -\frac{1}{\sqrt{2}}S_3^{(-1/3)} \epm,\label{eq:lqmultiplets}
\end{eqnarray}
with the superscripts referring to the electric charge of the leptoquark eigenstates. Expanding the multiplets in the Lagrangian of eq.~\eqref{eq:lag} in terms of the eigenstates of eq.~\eqref{eq:lqmultiplets}, we obtain:
\be\bsp
   \lag_{\rm LQ} = &\ \lag_{\rm kin.} + \lag_{\rm mass} + 
   {\bf y_{\sss 1}^{\sss RR}} \uRbar^c \eR S_1^{(+1/3)}
  +{\bf y_{\sss 1}^{\sss LL}} \big(\uLbar^c \eL S_1^{(+1/3)} - \dLbar^c \nuL S_1^{(+1/3)}\big)
 \\ &\ +
   {\bf y_{\sss 2}^{\sss LR}} \big(\eRbar \uL R_2^{(-5/3)} + \eRbar \dL R_2^{(-2/3)}\big)
  +{\bf y_{\sss 2}^{\sss RL}} \big(\uRbar \nuL R_2^{(+2/3)} - \uRbar \eL R_2^{(+5/3)}\big)
 \\ &\ +
   {\bf y_{\sss 3}^{\sss LL}} \big(\sqrt{2}\uLbar^c \nuL S_3^{(-2/3)} - \uLbar^c \eL S_3^{(+1/3)} - \dLbar^c \nuL S_3^{(+1/3)} - \sqrt{2}\dLbar^c \eL S_3^{(+4/3)}\big)
 \\ &\ +
   {\rm H.c.}
\esp\label{eq:lagexp}\ee

%%%%%%%%%%%%%%%%%%%%%
\subsection{Benchmark scenarios}
%%%%%%%%%%%%%%%%%%%%%
To illustrate the effect of $t$-channel diagrams, NLO-QCD corrections, and off-diagonal channels on scalar leptoquark pair production rates, we define below a few benchmark scenarios. We first consider generic simplified scenarios derived from models offering promising solutions to the flavour anomalies. Then we move on with an estimation of the impact of off-diagonal leptoquark production channels in the context of the phenomenologically motivated scenarios that were studied in depth in~\cite{Borschensky:2021hbo}.
 
%  (\textcolor{blue}{see also ref. \cite{Saad:2020ihm} for a similar study within a two-loop radiative neutrino mass model})
%%%%%%%%%%%%%%%%%%%%%%%%%%%%%%%%%%%%%%%%%%%%%%%
\subsubsection{General scenarios including $R_2$ and $S_3$ leptoquarks coupling to first- and second-generation quarks}\label{sec:benchmark_generic}
%%%%%%%%%%%%%%%%%%%%%%%%%%%%%%%%%%%%%%%%%%%%%%%
We begin the present work with a study of scalar leptoquark pair production in models involving simultaneously $R_2$ and $S_3$ leptoquarks, our calculations including both diagonal and off-diagonal channels. The term {\it diagonal} refers here to the production of a particle-antiparticle pair of the {\it same} leptoquark eigenstate, while the term {\it off-diagonal} stands for the production of two {\it different} leptoquark eigenstates. Consequently, in the case of a multi-species leptoquark model, the off-diagonal production mode does not only include the production of two different leptoquark states from a given multiplet, but also that of eigenstates of different multiplets.

To simplify the discussion and the physics analysis performed in our adopted $R_2$--$S_3$ model, we assume a mass degeneracy among all leptoquark states considered, and we compute the cross sections for two leptoquark masses\footnote{Whereas our lowest mass choice may be excluded by recent ATLAS results~\cite{ATLAS:2020dsk} that suggest a lower mass limit of 1.7--1.8 TeV on scalar leptoquarks coupling to muons and electrons, production characteristics are not drastically affected by the actual mass value in this range.},
\begin{eqnarray}
m_{\rm LQ} \equiv m_{R_2} = m_{S_3} = 1600~{\rm GeV}\quad\text{and}\quad 2400~{\rm GeV}.
\label{bs:masses}
\end{eqnarray}
In this scenario, off-diagonal leptoquark production includes seven channels. In order to make the discussion as generic as possible, we set all the relevant entries of the Yukawa coupling matrices to zero except $y_{2,12}^{\rm LR}$, $y_{2,12}^{\rm RL}$, and $y_{3,12}^{\rm LL}$. Furthermore, we assume a degeneracy among these couplings, and let them vary over the following range,
\begin{eqnarray}
y \equiv y_{2,12}^{\rm LR} = y_{2,12}^{\rm RL} = y_{3,12}^{\rm LL} \in [0.1, 1.5].
\label{bs:couplings}
\end{eqnarray}
Processes involving different leptoquark species hence depend on the product of two of the couplings of eq.~\eqref{bs:couplings}, which eventually leads to a $y^4$ dependence of the tree-level cross section. In appendix~\ref{app:amplitudes}, we detail how these different couplings lead to the opening of the different off-diagonal production modes, and how the associated cross sections are related to each other.

In another scenario, we ignore the contributions of the $S_3$ leptoquark and analyse a model containing only $R_2$ leptoquarks. We consider a dependence on the Yukawa couplings in two configurations: {\it(i)} couplings to first-generation quarks only and {\it(ii)} couplings to second-generation quarks only,
\begin{equation}
\begin{split}
\text{\it(i): } y_{2,12}^{\rm LR} \in [0.1, 1.5], \qquad y_{2,22}^{\rm LR} = 0,\\
\text{\it(ii): } y_{2,22}^{\rm LR} \in [0.1, 1.5], \qquad y_{2,12}^{\rm LR} = 0.
\end{split}
\end{equation}
We set in both cases all other couplings to zero, and in particular $y_{2,12}^{\rm RL} = y_{2,22}^{\rm RL} = 0$.

%%%%%%%%%%%%%%%%%%%%%%
\subsubsection{Benchmark scenarios addressing the flavour anomalies}\label{sec:benchmarksFlavor}
%%%%%%%%%%%%%%%%%%%%%%

\paragraph{Phenomenologically-viable $R_2$ models\! $-$} \label{sec:benchmr2}

\renewcommand{\arraystretch}{1.35}\setlength\tabcolsep{10pt}
\begin{table}
 \centering
 \begin{tabular}{l | c | ccc}
   & $y^{\rm RL}_{2,23}$ & $y^{\rm LR}_{2,33}$ & $y^{\rm LR}_{2,21}$ & $y^{\rm LR}_{2,31}$
  \tabularnewline
  \hline
  $a_1$ & $1.84+1.84i$ & $0.354+0.354i$ & $-0.015i$ & $0.262+0.262i$
  \tabularnewline
  $a_2$ & $0.309+0.951i$ & $0.951+0.309i$ & $0.011-0.011i$ & $0.37i$
  \tabularnewline
 \end{tabular}
 \caption{Non-zero Yukawa couplings for two benchmark points in a model that includes a single $R_2$ leptoquark species (with a mass of 1 TeV), and that provides an explanation for the flavour anomalies according to~\cite{Popov:2019tyc}. The two points have been selected from the $1\sigma$ ranges of model parameters allowing to explain the anomalies.}
 \label{tab:benchmarksR2}
\end{table}

\hspace{-.3cm}An interesting scenario that minimally addresses all flavour anomalies with a unique species of scalar leptoquarks has been suggested in \cite{Popov:2019tyc}. In that scenario, the SM is extended by a single leptoquark doublet $R_2$ that couples to tau leptons and to electrons, which is sufficient to address both the $R_{D^{(*)}}$ and the $R_{K^{(*)}}$ anomalies. In this context and based on the results of \cite{Popov:2019tyc}, we consider two benchmark points denoted by $a_1$ and $a_2$ with $m_{R_2} = 1000~{\rm GeV}$ (see table \ref{tab:benchmarksR2}). Whereas these choices lead to mild tensions between data and theory for the ${\rm BR}(B_c\to\tau\nu)$ branching ratio, these can be weakened by considering non-minimal models involving two leptoquark species, like $R_2$ and $S_3$.

\paragraph{A two-leptoquark $R_2$--$S_3$ model inspired by Grand Unification \!$-$}\label{sec:benchmr2s3}

\renewcommand{\arraystretch}{1.35}\setlength\tabcolsep{7pt}
\begin{table}
 \centering
 \begin{tabular}{l | c | cc | cccc }
  & $y_{2,33}^{\rm LR}$ & $y_{2,22}^{\rm RL}$ & $y_{2,23}^{\rm RL}$ & $y_{3,22}^{\rm LL}$ & $y_{3,23}^{\rm LL}$ & $y_{3,32}^{\rm LL}$ & $y_{3,33}^{\rm LL}$
  \tabularnewline
  \hline
  $b_1$ & $-0.18734+1.12287i$ & $0.265001$ & $1.17382$ & $-0.010$ & $-0.045$ & $-0.265$ & $-1.173$
  \tabularnewline
  $b_2$ & $-0.18734+1.12287i$ & $0.37353$ & $1.59511$ & $-0.014$ & $-0.061$ & $-0.373$ & $-1.594$
  \tabularnewline
 \end{tabular}
 \caption{Non-zero Yukawa couplings for two benchmark points in a two-leptoquark model allowing for an explanation for the flavour anomalies~\cite{Becirevic:2018afm,Becirevic:2020rzi}. Benchmarks include an $R_2$ leptoquark (with $m_{R_2} = 1.3$\,TeV) and an $S_3$ leptoquark (with $m_{S_3} = 2$\,TeV). Point $b_1$ corresponds to the best fit value and $b_2$ is chosen inside the $2\sigma$ region returned by the fit.}
 \label{tab:benchmarksR2plusS3}
\end{table}

\hspace{-.3cm}Here, the SM is extended by several leptoquark species with masses lying in the TeV regime. Such a model was proposed in~\cite{Becirevic:2018afm}, where it was shown that the presence of two leptoquarks $R_2$ and $S_3$ suffices to address the $R_{D^{(*)}}$ and $R_{K^{(*)}}$ anomalies and to avoid all constraints from direct and indirect searches for leptoquarks at the LHC, from LEP-I precision measurements at the $Z$-pole, and from measurements of other flavour observables. Following the updated fit of~\cite{Becirevic:2020rzi}, we consider two scenarios dubbed $b_1$ and $b_2$ with $m_{R_2} = 1300$~GeV and $m_{S_3} = 2000$~GeV (see table \ref{tab:benchmarksR2plusS3})\footnote{Similar models have been studied in refs. \cite{Saad:2020ihm,Saad:2020ucl} within a two-loop radiative neutrino mass  generation.}. 

\paragraph{The singlet-triplet $S_1$--$S_3$ leptoquark model\! $-$}\label{sec:benchms1s3}

\renewcommand{\arraystretch}{1.35}\setlength\tabcolsep{4.5pt}
\begin{table}
\centering
 \begin{tabular}{l | cccc | cc | cccc}
    & $y_{1,22}^{\rm LL}$ & $y_{1,23}^{\rm LL}$ & $y_{1,32}^{\rm LL}$ & $y_{1,33}^{\rm LL}$ & $y_{1,23}^{\rm RR}$ & $y_{1,32}^{\rm RR}$ & $y_{3,22}^{\rm LL}$ & $y_{3,23}^{\rm LL}$ & $y_{3,32}^{\rm LL}$ & $y_{3,33}^{\rm LL}$
  \tabularnewline
  \hline
  $c_1$ & $-0.0082$ & $-1.46$ & $-0.016$ & $-0.064$ & $1.34$ & $-0.19$ & $-0.019$ & $0.58$ & $-0.059$ & $-0.11$
  \tabularnewline
  $c_2$ & $0.0078$ & $1.36$ & $-0.055$ & $0.052$ & $-1.47$ & $-0.053$ & $-0.017$ & $-1.23$ & $-0.070$ & $0.066$
  \tabularnewline
 \end{tabular}
 \caption{Non-zero Yukawa couplings for two benchmark points in a two-leptoquark scenario providing an explanation for the flavour anomalies~\cite{Crivellin:2019dwb}. Benchmarks include $S_1$and $S_3$ leptoquarks (with $\mlq = 1.2$\,TeV). Points $c_1$ and $c_2$ correspond to the $p_1$ and $p_2$ benchmarks in the notation of~\cite{Crivellin:2019dwb}. }
 \label{tab:benchmarksS1plusS3}
\end{table}

\hspace{-.3cm}Ref.~\cite{Crivellin:2019dwb} demonstrated that a possible solution to the flavour anomalies could be designed by extending the SM by two leptoquark species $S_1$ and $S_3$, provided that the leptoquarks couple to muons and tau leptons. In addition, this framework reduces the gap between the theory prediction and the experimental measurement of the anomalous magnetic moment of the muon. We use two benchmark points $c_1$ and $c_2$, that correspond to the $p_1$ and $p_2$ setups from~\cite{Crivellin:2019dwb}, in which we choose leptoquark masses $m_{S_1} = m_{S_3} = 1200$~TeV. Details about these benchmarks can be found in table~\ref{tab:benchmarksS1plusS3}.

%%%%%%%%%%%%%%%%%%%%%%%%%%%%%
\subsection{Leptoquark pair production at next-to-leading order}
%%%%%%%%%%%%%%%%%%%%%%%%%%%%%

\begin{figure}
\centering
\includegraphics[width=0.9\linewidth]{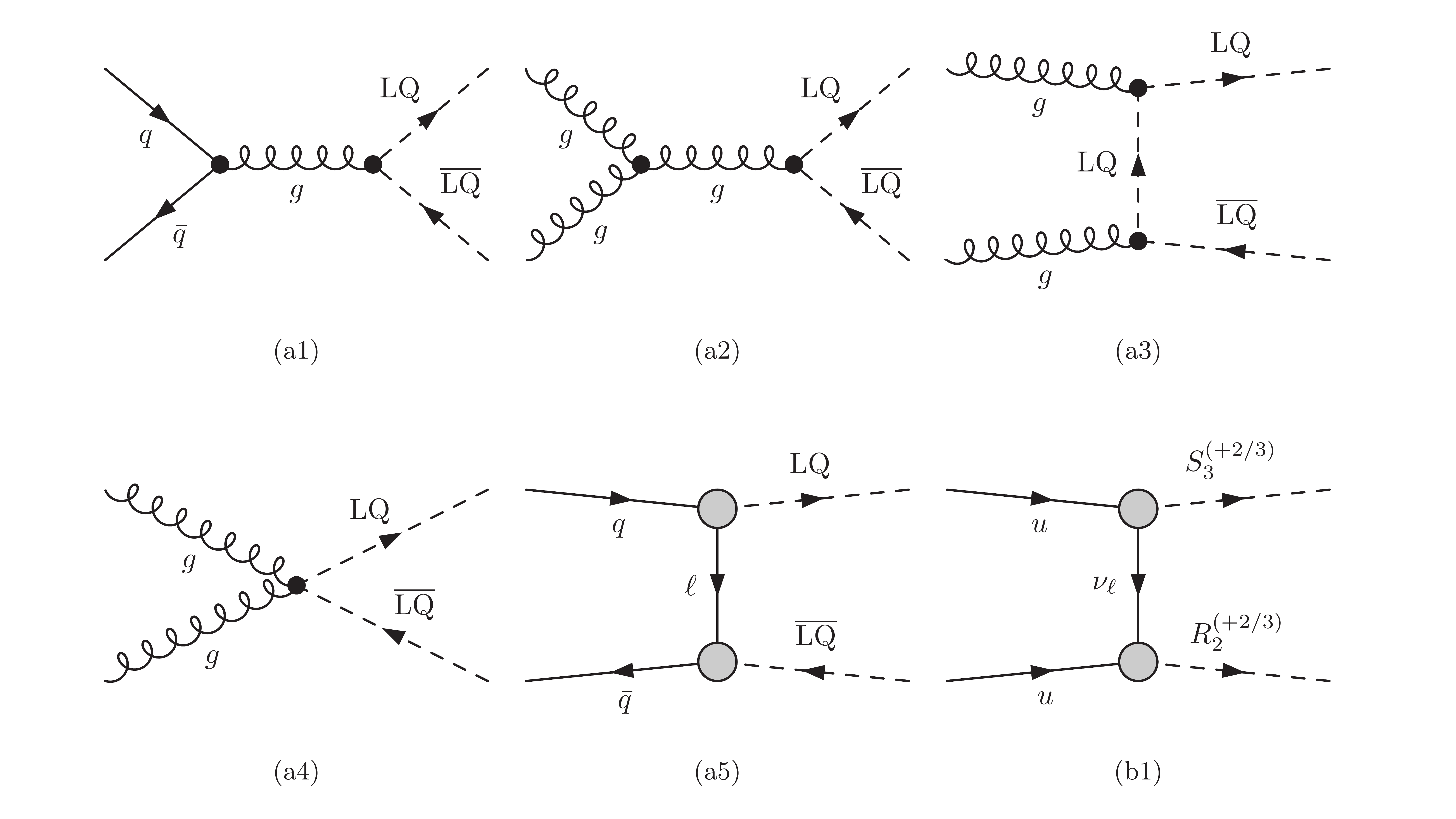} 
\includegraphics[width=0.9\linewidth]{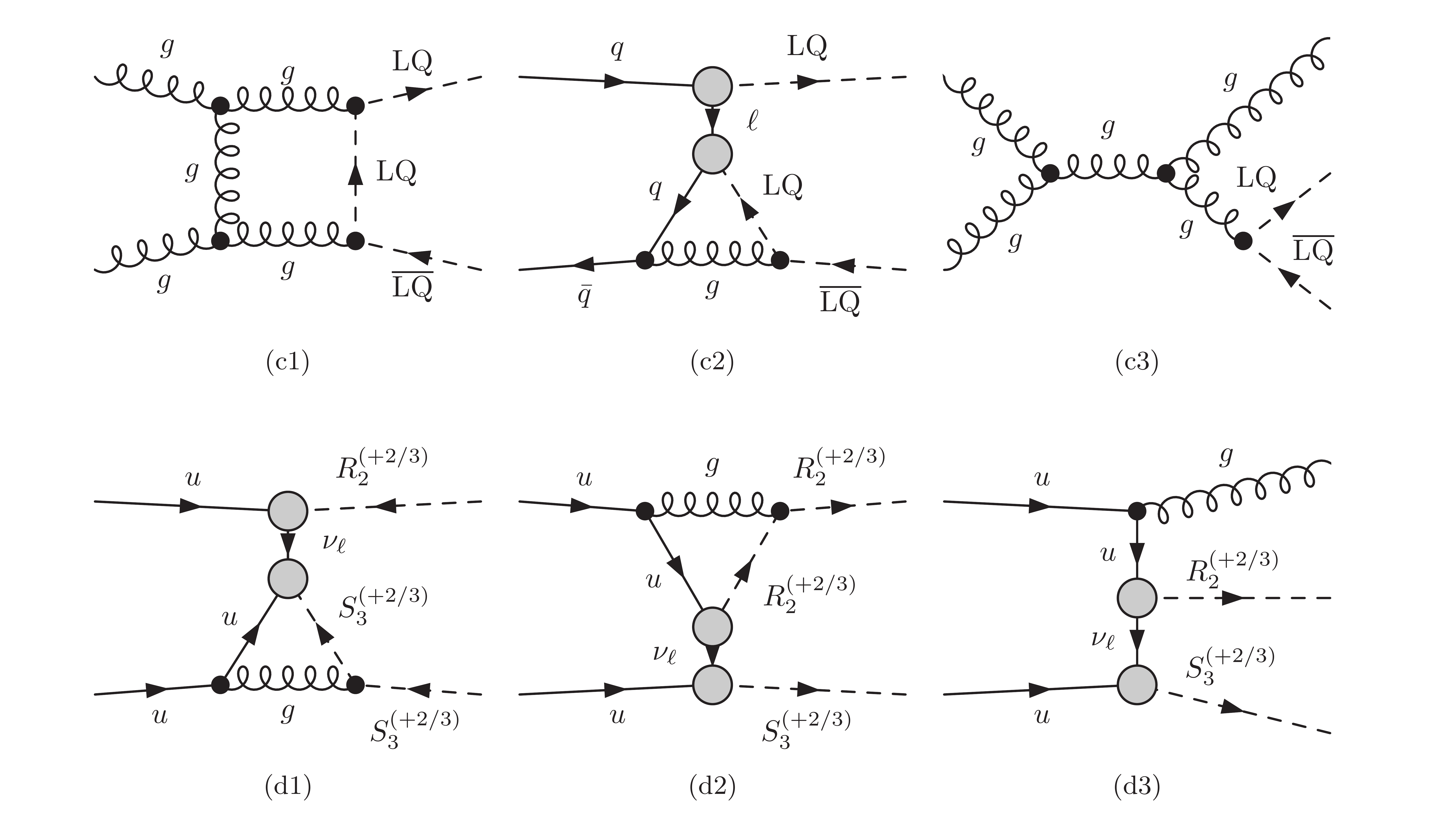}
\caption{Selection of Feynman diagrams contributing to scalar leptoquark pair production in hadron collisions, both for the diagonal and off-diagonal channel (for which we have chosen the $pp\to R_2^{(+2/3)} S_3^{(+2/3)}$ example). QCD couplings are indicated by a black point, while a grey circle denotes the lepton-quark-leptoquark Yukawa couplings. Born diagrams are shown in the sub-panels (a1)--(a5) for the diagonal case, and (b1) for the off-diagonal one. Virtual correction and real emission diagrams are shown in sub-panels (c1)--(c3) and (d1)--(d3) respectively.}
\label{fig:FD}
\end{figure}

We discuss in this section general features of scalar leptoquark pair production in proton-proton collisions. We consider the following process
\begin{eqnarray}
p p \to {\rm LQ}_i\, {\rm LQ}_j + X,
\end{eqnarray}
where ${\rm LQ}_i$ and ${\rm LQ}_j$ refer generically to any (possibly identical) leptoquark species. In the following, we always sum over charge-conjugate processes when they are different. For instance, $p p \to S_3^{(\pm1/3)} S_3^{(\pm 2/3)}$ would refer to both the production of an $S_3^{(+1/3)} S_3^{(+2/3)}$ and an $S_3^{(-1/3)} S_3^{(-2/3)}$ pair. The corresponding NLO total production rate is given by
\be
\sigma_{\rm NLO} = \sum_{a,b=q, \bar q, g} \int {\rm d} x_a\ {\rm d} x_b\ f_{a/p} (x_a, \mu_F^2)\ f_{b/p} (x_b, \mu_F^2)\ \hat{\sigma}_{\rm NLO}(ab\to {\rm LQ}_i\, {\rm LQ}_j),
\label{eq:factorisation}
\ee
where $x_i$ stands for the momentum fraction of the parton $i$ inside the proton, $f_{i/p}$ for its parton distribution function evaluated at a factorisation scale $\mu_F$, and $\hat{\sigma}$ is the (NLO-QCD) partonic cross section.  It includes Born contributions $\sigma^{(0)}$ and ${\cal O}(\alpha_s)$ corrections $\sigma^{(1)}$ for both of which a selection of Feynman diagrams is shown in figure~\ref{fig:FD}, in the diagonal and off-diagonal case. 

For the diagonal channel, the Born component of the cross section receives pure QCD contributions that are proportional to $\alpha_s^2$ [diagrams (a1)--(a4)], $t$-channel lepton exchange contributions that are proportional to $y^4$ [diagram (a5)], and the interference between the two that is of ${\cal O}(\alpha_s y^2)$. In contrast, only $t$-channel exchange diagrams survive for off-diagonal leptoquark pair production modes [diagram (b1)]. NLO corrections to the diagonal case therefore include three classes of contributions, namely one proportional to $\alpha_s^3$ (the pure QCD component of $\sigma^{(1)}$), one proportional to $\alpha_s y^4$ (NLO-QCD corrections to $t$-channel exchange contributions) and one proportional to $\alpha_s^2 y^2$ (NLO-QCD corrections to the interference between QCD and $t$-channel amplitudes). In the off-diagonal case, only ${\cal O}(\alpha_s y^4)$ contributions are obviously relevant. Correspondingly, diagrams (c1)--(c2) and (d1)--(d2) illustrate virtual loop corrections, whereas diagrams (c3) and (d3) are examples of real-emission corrections.

%%%%%%%%%%%%%%%%%%%%%%%%%%%%%%%%%%%%%%%%%%%%%%%%%%
\section{Scalar leptoquark pair production in general scenarios}
\label{sec:results}
%%%%%%%%%%%%%%%%%%%%%%%%%%%%%%%%%%%%%%%%%%%%%%%%%%

\subsection{Technical setup}
Calculations of fixed-order cross sections at LO and NLO are performed using \textsc{MadGraph\_aMC@NLO} version 3.1.1 \cite{Alwall:2014hca} and the \texttt{LQnlo\_5FNS\_v5} UFO~\cite{Degrande:2011ua} model documented in~\cite{Borschensky:2020hot,Borschensky:2021hbo}\footnote{The model is available from \url{https://www.uni-muenster.de/Physik.TP/research/kulesza/leptoquarks.html}.}. To assess the dependence of the results on the PDF choice, we use two LO and three NLO PDF sets accessed through \textsc{Lhapdf} 6.4.0~\cite{Buckley:2014ana}: \texttt{MSHT20lo\_as\_130}~\cite{Bailey:2020ooq} (with $\alpha_s(M_Z) = 0.130$), \texttt{NNPDF40\_lo\_as\_01180}~\cite{Ball:2021leu}, \texttt{MSHT20nlo\_as\_118}~\cite{Bailey:2020ooq}, \texttt{CT18NLO}~\cite{Hou:2019efy} and \texttt{NNPDF40\_nlo\_as\_01180}~\cite{Ball:2021leu} (all four last sets using $\alpha_s(M_Z) = 0.118$). Moreover, we fix the renormalisation and the factorisation scale to the mass of the produced scalar leptoquarks $\mu_R = \mu_F = m_{\rm LQ}$, and estimate scale uncertainties with the seven-point method (in which the error is computed from the envelope spanned by the cross section when varying the scales independently by a factor of 2 up and down, while keeping their ratio to at most 2).

In order to validate our predictions, we compute independently cross sections using the \pwb{} v2 framework~\cite{Nason:2004rx,Frixione:2007vw,Alioli:2010xd}. Here, the relevant virtual amplitudes have been generated with \textsc{FeynArts}~\cite{Hahn:2000kx} and \textsc{FormCalc}~\cite{Hahn:1998yk}, and are based on a manual implementation of the leptoquark interaction vertices in an {\it ad hoc} \textsc{FeynArts} model. We have found excellent agreement between the \mgamc{} and \pwb{} predictions.

Technical details about the necessary modifications of the \mgamc{} and \pwb{} packages to compute leptoquark production rates and about how to run the codes can be found in~\cite{Borschensky:2021hbo}.

%%%%%%%%%%%%%%%%%%%%%%%%%%%%%%%%%%%%%%%%%%%
\subsection{NLO cross sections in the $R_2$--$S_3$ model}
%%%%%%%%%%%%%%%%%%%%%%%%%%%%%%%%%%%%%%%%%%%

\begin{figure}
    \centering
    \includegraphics[width=0.85\linewidth]{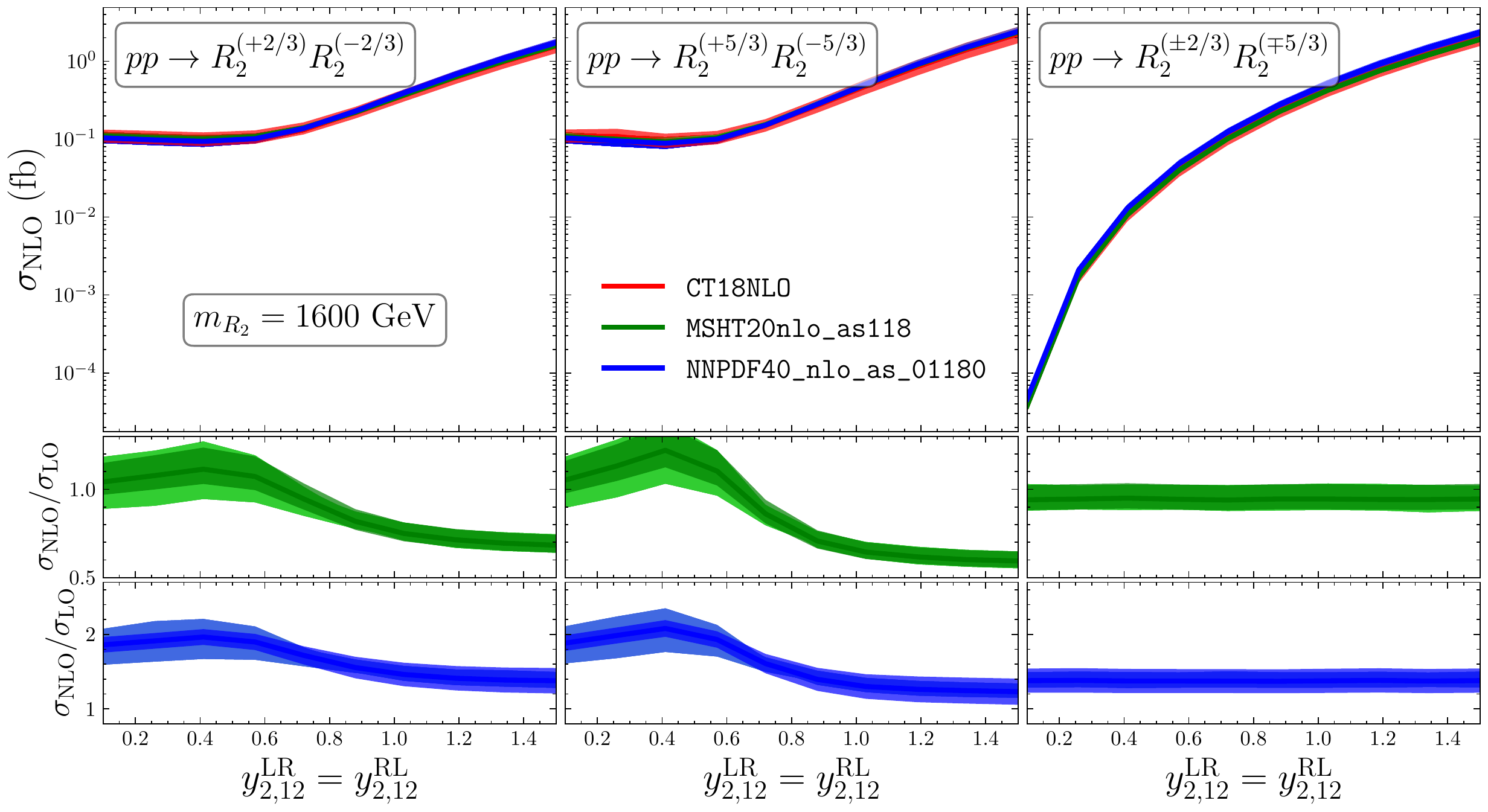}
    \caption{Total cross sections for the production of a pair of $R_2$ eigenstates as a function of $y = y_{2,12}^{\rm LR} = y_{2,12}^{\rm RL}$ for the three processes $pp\to R_2^{(+2/3)} R_2^{(-2/3)}$ (left), $pp\to R_2^{(+5/3)} R_2^{(-5/3)}$ (middle) and $pp\to R_2^{(\pm 5/3)} R_2^{(\mp 2/3)}$ (right). In the lower panels, we present NLO $K$-factors defined as the ratio of the NLO rates to the LO ones for the MSHT20 and CT18 PDF sets. Our predictions include scale (light error bands) and PDF (darker error bands) uncertainties at NLO (evaluated relatively to the central LO predictions for the $K$-factors).}
    \label{fig:xsec:R2R2:CH1:y12:MLQ:1600}
\end{figure}

We focus first in figure~\ref{fig:xsec:R2R2:CH1:y12:MLQ:1600} on total cross sections for the production of leptoquarks of the $R_2$ species, the leptoquark mass being fixed to 1600~GeV. While the plots on the left and in the middle panels of the figure present results for the two diagonal production modes, \ie\ for the production of $R_2^{(+2/3)}R_2^{(-2/3)}$ and $R_2^{(+5/3)}R_2^{(-5/3)}$ leptoquark pairs, the plot on the right of the figure concerns the off-diagonal production of an $R_2^{(\pm 2/3)}R_2^{(\mp 5/3)}$ pair of leptoquarks (the contributions of the two conjugate processes being summed over). In the former case, the leptoquark Yukawa coupling $y$ impacts the NLO rates only for values larger than $y\sim0.6$, the pure QCD channel being largely dominant for smaller $y$ values. In contrast, the off-diagonal channel only involves diagrams depending on the Yukawa coupling $y$, so that the corresponding cross section is negligible for small values of $y$ and grows steadily with the coupling. At larger values of $y$, it delivers a comparable contribution to that of the diagonal mode with $t$-channel exchange included. 

In addition, we compare total rate predictions obtained with the \texttt{MSHT20nlo\_as\_118} (green), \texttt{CT18NLO} (red) and \texttt{NNPDF40\_nlo\_as\_01180} (blue) sets of parton densities in the upper panel of the figure. In the lower panel we show $K$-factors defined as the ratio of the NLO rates to the LO ones when MSHT20 and NNPDF4.0 densities are used, LO rates being respectively evaluated with the \texttt{MSHT20lo\_as\_130} and \texttt{NNPDF40\_lo\_as\_01180} PDF sets\footnote{$K$-factors associated with CT18 densities have not been computed as the CT18 LO set is not publicly available.}. Whereas all NLO cross sections are found to agree within their uncertainties, the values of the $K$-factors strongly depend on the used PDF set, as visible from the lower panels of figure~\ref{fig:xsec:R2R2:CH1:y12:MLQ:1600}. This behaviour originates from strong differences inherent to the LO PDF sets, which emphasises that NLO predictions are mandatory for reliable results. Moreover, for all channels and all choices of parton densities, NLO corrections are significant. Their effects are found to modify LO rates by a value ranging from tens of percents to up to a factor of 2, the exact value depending on $y$ and the PDF set.

The shape of the $K$-factors as a function of the Yukawa coupling $y$ is found to be similar for all parton densities considered, both for the diagonal and the off-diagonal channels. As expected, there is no dependence of the NLO-QCD $K$-factor on the value of the Yukawa coupling in the off-diagonal case, as the $y^4$ dependence of the cross section cancels out in the ratio. Conversely, in the diagonal production modes the $K$ factor dependence on $y$ clearly depicts the transition (around $y\sim 0.6$) between the region of the parameter space in which the cross section is dominated by QCD diagrams, and that in which it is dominated by $t$-channel diagrams.

\begin{figure}
    \centering
    \includegraphics[width=0.85\linewidth]{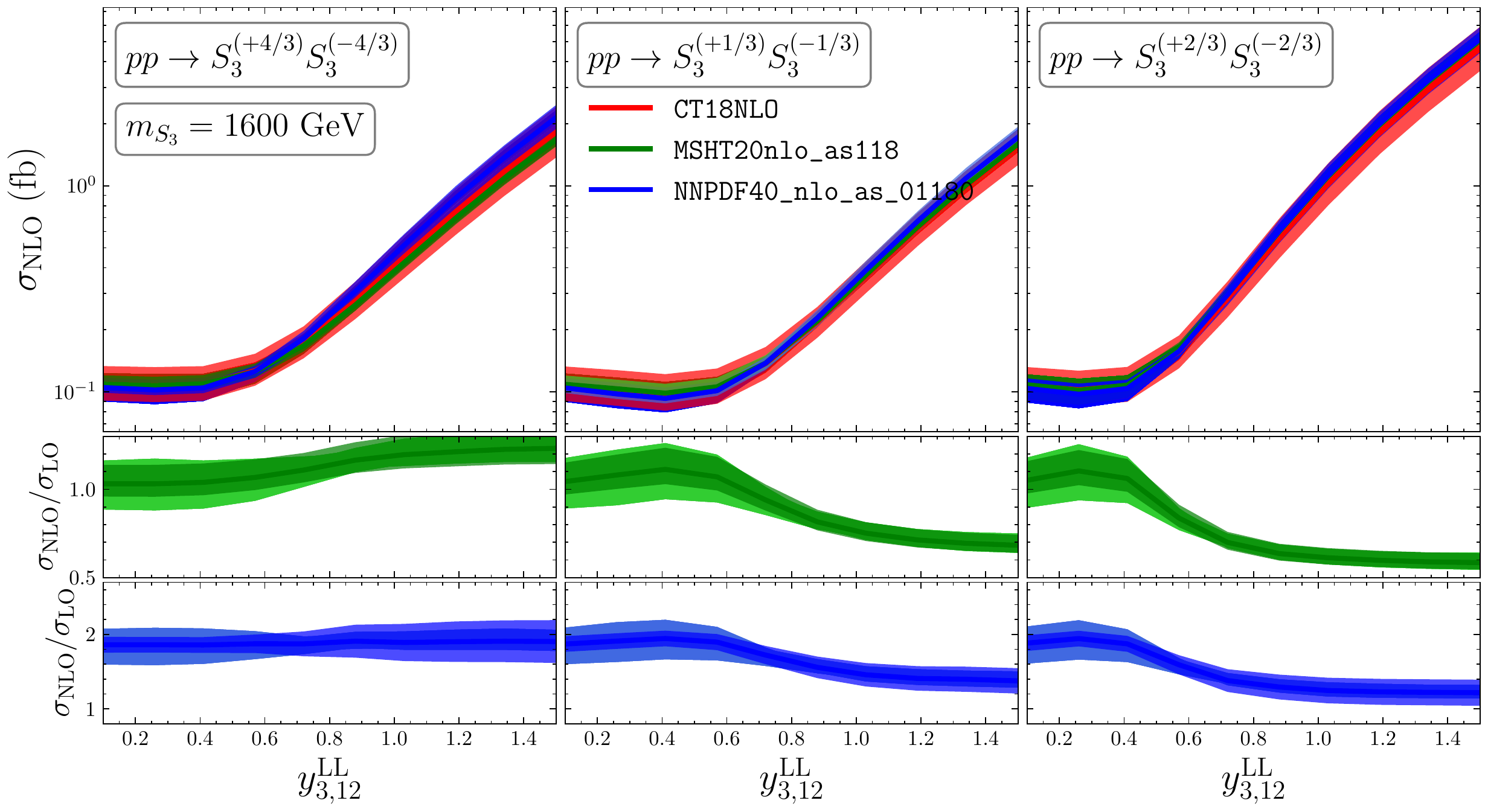}
    \vfill
    \includegraphics[width=0.6\linewidth]{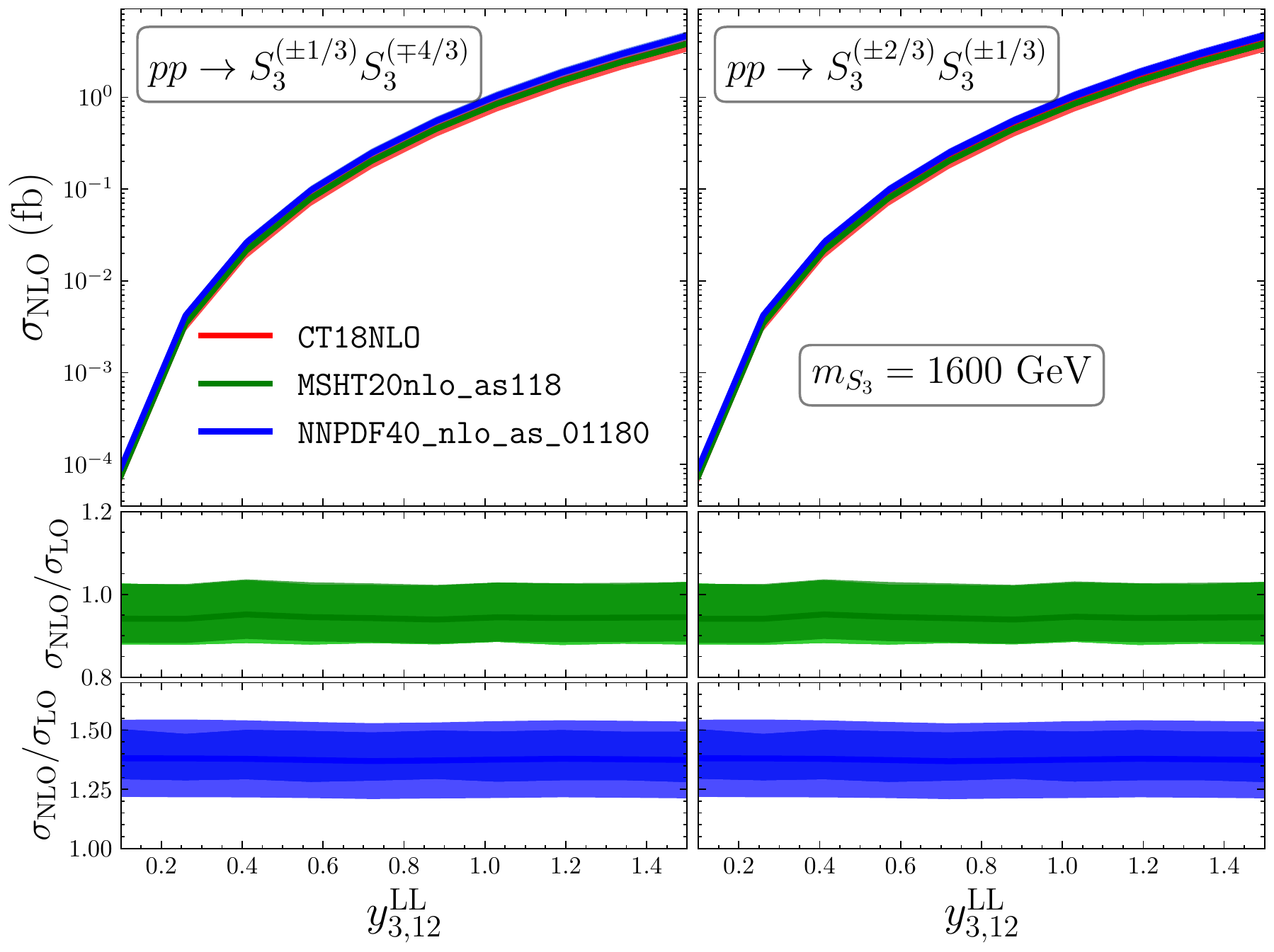}
    \caption{Same as in figure \ref{fig:xsec:R2R2:CH1:y12:MLQ:1600} but for processes involving solely $S_3$ leptoquark eigenstates. In the upper panels, we shown results as a function of $y=y_{3,12}^{\rm LL}$ for the diagonal channels, \ie\ $p p \to S_3^{(+4/3)} S_3^{(-4/3)}$ (left), $p p \to S_3^{(+1/3)} S_3^{(-1/3)}$ (middle) and $p p \to S_3^{(+2/3)} S_3^{(-2/3)}$ (right), whereas in the lower panels we focus on the off-diagonal production modes $pp\to S_3^{(\pm 1/3)} S_3^{(\mp 4/3)}$ (left) and $p p\to S_3^{(\pm 2/3)} S_3^{(\pm 1/3)}$ (right).}
    \label{fig:xsec:S3S3:CH2:y12:MLQ:1600}
\end{figure}

In figure~\ref{fig:xsec:S3S3:CH2:y12:MLQ:1600} we consider the production of a pair of $S_3$ leptoquark eigenstates, the leptoquark mass being fixed again to 1600~GeV. In the top row of the figure, we study the dependence of the cross sections and the $K$-factors on the Yukawa coupling $y =y_{3,12}^{\rm LL}$ for the three diagonal processes ($p p \to S_3^{(+4/3)} S_3^{(-4/3)}$, $p p \to S_3^{(+1/3)} S_3^{(-1/3)}$ and $p p \to S_3^{(+2/3)} S_3^{(-2/3)}$), whereas the bottom row of the figure is dedicated to predictions for the off-diagonal production modes ($pp\to S_3^{(\pm 1/3)} S_3^{(\mp 4/3)}$ and $p p\to S_3^{(\pm 2/3)} S_3^{(\pm 1/3)}$). Cross sections for the production of a diagonal pair of $S_3$ leptoquark states feature characteristics that are very similar to those associated with the diagonal production of a pair of $R_2$ leptoquark eigenstates, which we studied in figure~\ref{fig:xsec:R2R2:CH1:y12:MLQ:1600}. NLO predictions indeed grow with increasing Yukawa coupling values due to $t$-channel contributions that are more and more important and that significantly impact rates once they become non-negligible relative to QCD contributions. In addition, NLO predictions obtained with the three PDF sets considered agree within their errors. 

\begin{figure}
    \centering
    \includegraphics[width=1.02\linewidth]{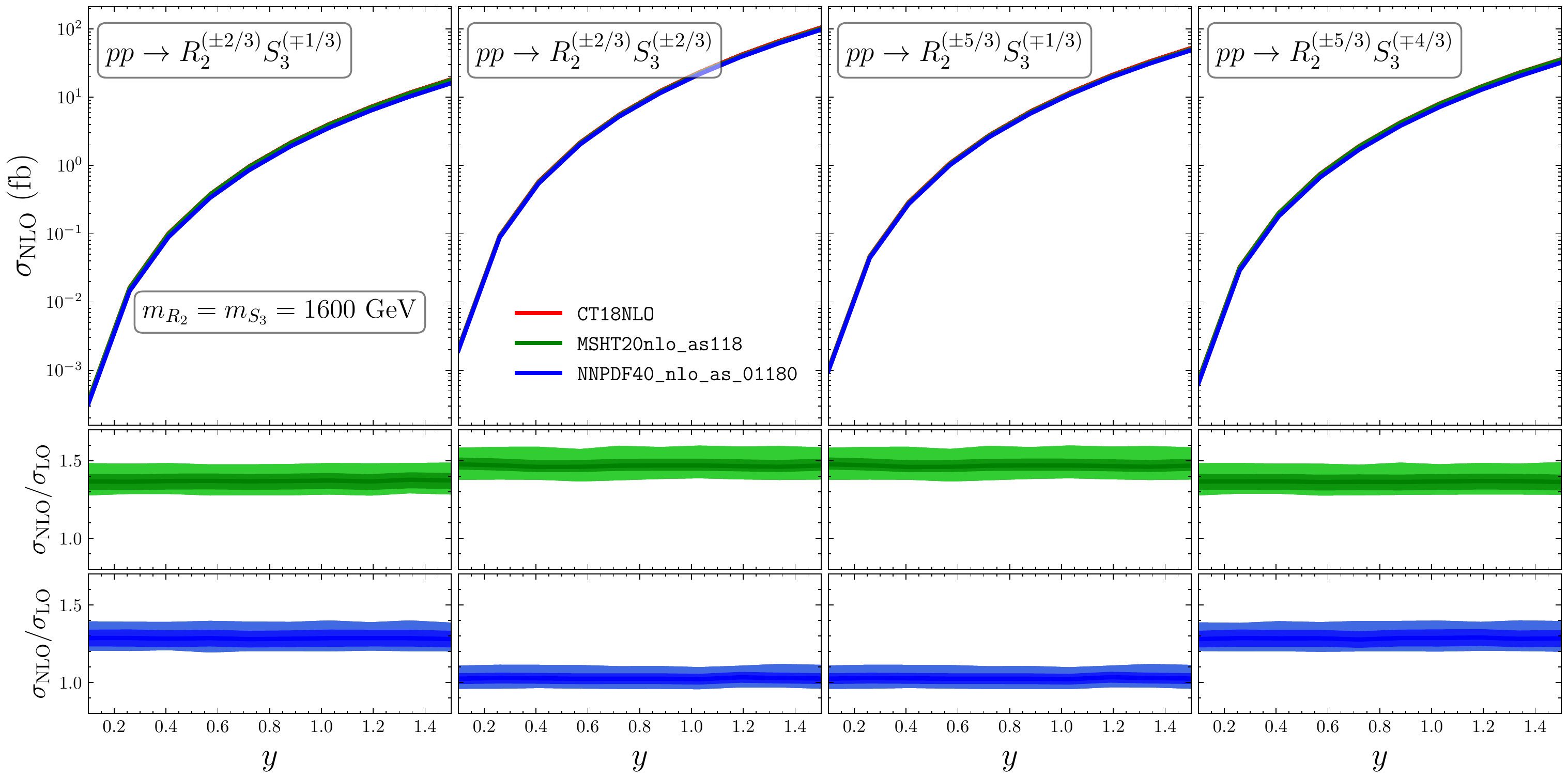}
    \vspace{-0.5cm}
    \caption{Same as in figure \ref{fig:xsec:R2R2:CH1:y12:MLQ:1600} but for the off-diagonal processes $pp\to R_2^{(\pm2/3)} S_3^{(\mp 1/3)}$ (first column), $pp\to R_2^{(\pm2/3)} S_3^{(\pm2/3)}$ (second column), $pp\to R_2^{(\pm5/3)} S_3^{(\mp1/3)}$ (third column) and $pp\to R_2^{(\pm5/3)} S_3^{(\mp4/3)}$ (fourth column). Cross sections and $K$-factors are presented as a function of the leptoquark Yukawa coupling $y=y_{2,12}^{\rm LR}=y_{3,12}^{\rm LL}$.}
    \label{fig:xsec:R2S3:CH1:y12:MLQ:1600}
\end{figure}

For the $pp\to S_3^{(+2/3)} S_3^{(-2/3)}$ and $pp\to S_3^{(+1/3)} S_3^{(-1/3)}$ processes, NLO K-factors are found to exhibit a similar behaviour as for the $R_2$ diagonal production modes, with a change of behaviour at $y\sim 0.4$--$0.6$ which corresponds to the threshold at which $t$-channel diagrams start to contribute. For the $pp\to S_3^{(+4/3)} S_3^{(-4/3)}$ process, the Yukawa coupling value at which $t$-channel diagrams start to contribute is similar ($y\sim 0.4$--$0.6)$, but the $K$-factor value increases, instead of decreases. This different behaviour originates from the related $t$-channel contributions that are solely $d\bar d$ initiated, unlike for the other two diagonal processes for which up-type quarks contribute. For all three diagonal processes, $K$-factors are found to affect LO cross sections by a few dozens of percent to up to a factor of 2, the exact value depending on the chosen PDF sets.

We now consider the off-diagonal processes $pp\to S_3^{(\pm 1/3)} S_3^{(\mp 4/3)}$ as well as $p p\to S_3^{(\pm 2/3)} S_3^{(\pm 1/3)}$, which we discuss together with processes in which one eigenstate from the $R_2$ multiplet is produced in association with one eigenstate from the $S_3$ multiplet, $pp\to R_2^{(\pm2/3)} S_3^{(\mp 1/3)}$, $pp\to R_2^{(\pm2/3)} S_3^{(\pm2/3)}$, $pp\to R_2^{(\pm5/3)} S_3^{(\mp1/3)}$, and $pp\to R_2^{(\pm5/3)} S_3^{(\mp4/3)}$. Cross sections and $K$-factors are respectively shown in the bottom row of figure~\ref{fig:xsec:S3S3:CH2:y12:MLQ:1600}, and in figure~\ref{fig:xsec:R2S3:CH1:y12:MLQ:1600}. Qualitatively, they all present features that are very similar to those arising in $R_2^{(\pm2/3)}R_2^{(\mp5/3)}$ production.  Cross section predictions at NLO are independent of the parton density choice (within uncertainties), and the $K$-factors are constant as the Yukawa coupling dependence in $y^4$ cancels out in the NLO to LO cross section ratios. We refer to appendix~\ref{app:amplitudes} for general details on the cross section dependence on the different Yukawa couplings entering the Lagrangian~\eqref{eq:lag}, that we take all equal here for simplicity. However, the obtained values for the $K$-factors associated with the various processes depend on the chosen PDF set, which is due to differences inherent to the \texttt{MSHT20lo\_as\_130} and \texttt{NNPDF40\_lo\_as\_01180} LO sets.

Crucially, at large values of the Yukawa coupling $y$, the NLO cross sections are substantially higher for processes involving leptoquarks of two different species than for any of the other processes considered here (both the diagonal channels and the production of a pair of different leptoquark eigenstates of the same species). The reason stems from the necessity of producing such a final state, that carries a total fermion number ($F=3B+L$, with $B$ and $L$ being the baryon and lepton quantum numbers) equal to 2, from a pair of valence quarks~\cite{Dorsner:2021chv}.

%%%%%%%%%%%%%%%%%%%%%%%%%%%%%%%%%%%%%%%%%%%%%%%%
\subsection{$t$-channel and off-diagonal contributions on inclusive leptoquark production rates in the $R_2$--$S_3$ model}
\label{sec:tchannel:MLQ:1600}
%%%%%%%%%%%%%%%%%%%%%%%%%%%%%%%%%%%%%%%%%%%%%%%%

\begin{figure}
    \centering
    \includegraphics[width=0.48\linewidth]{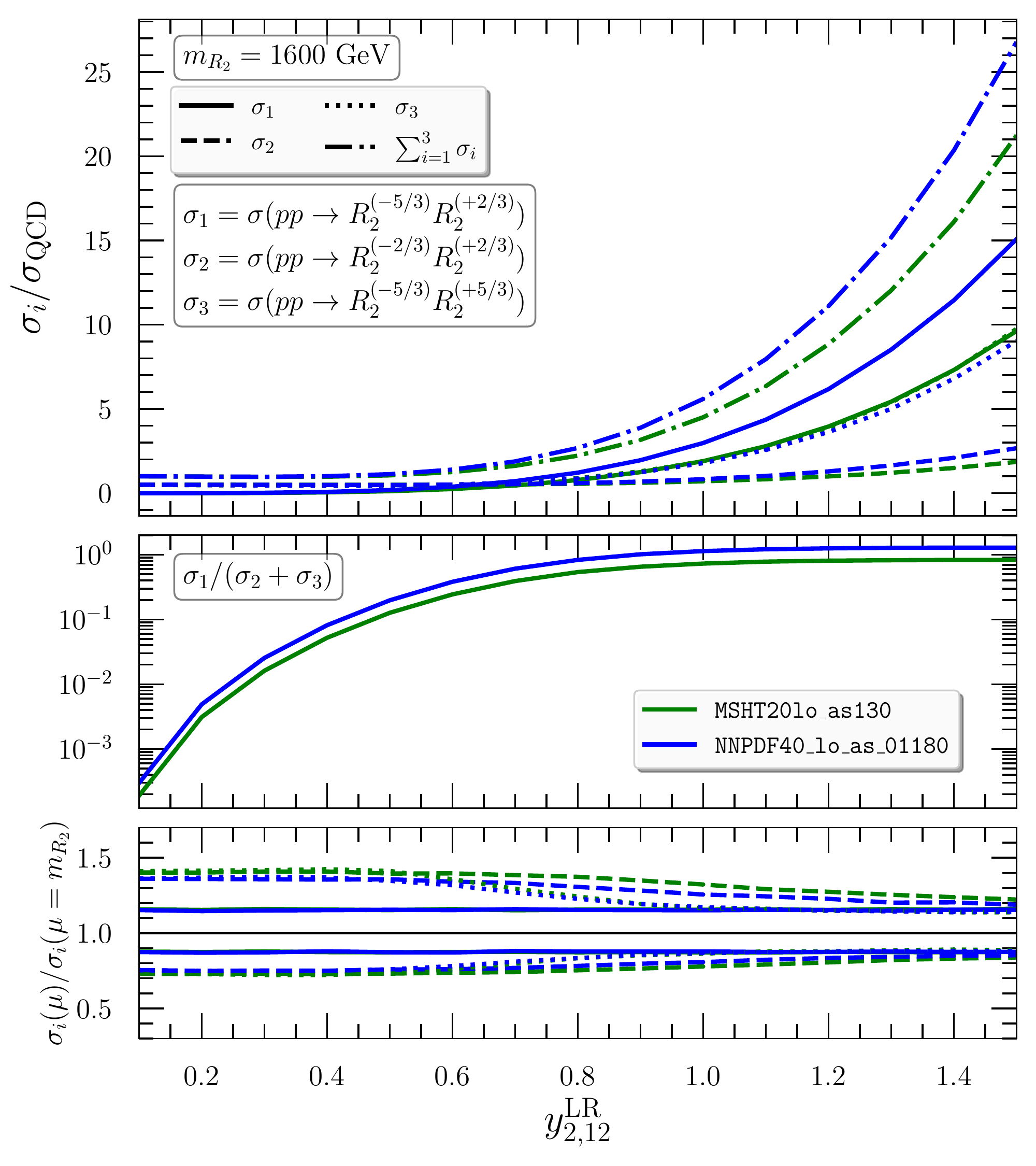}
    \hfill
    \includegraphics[width=0.48\linewidth]{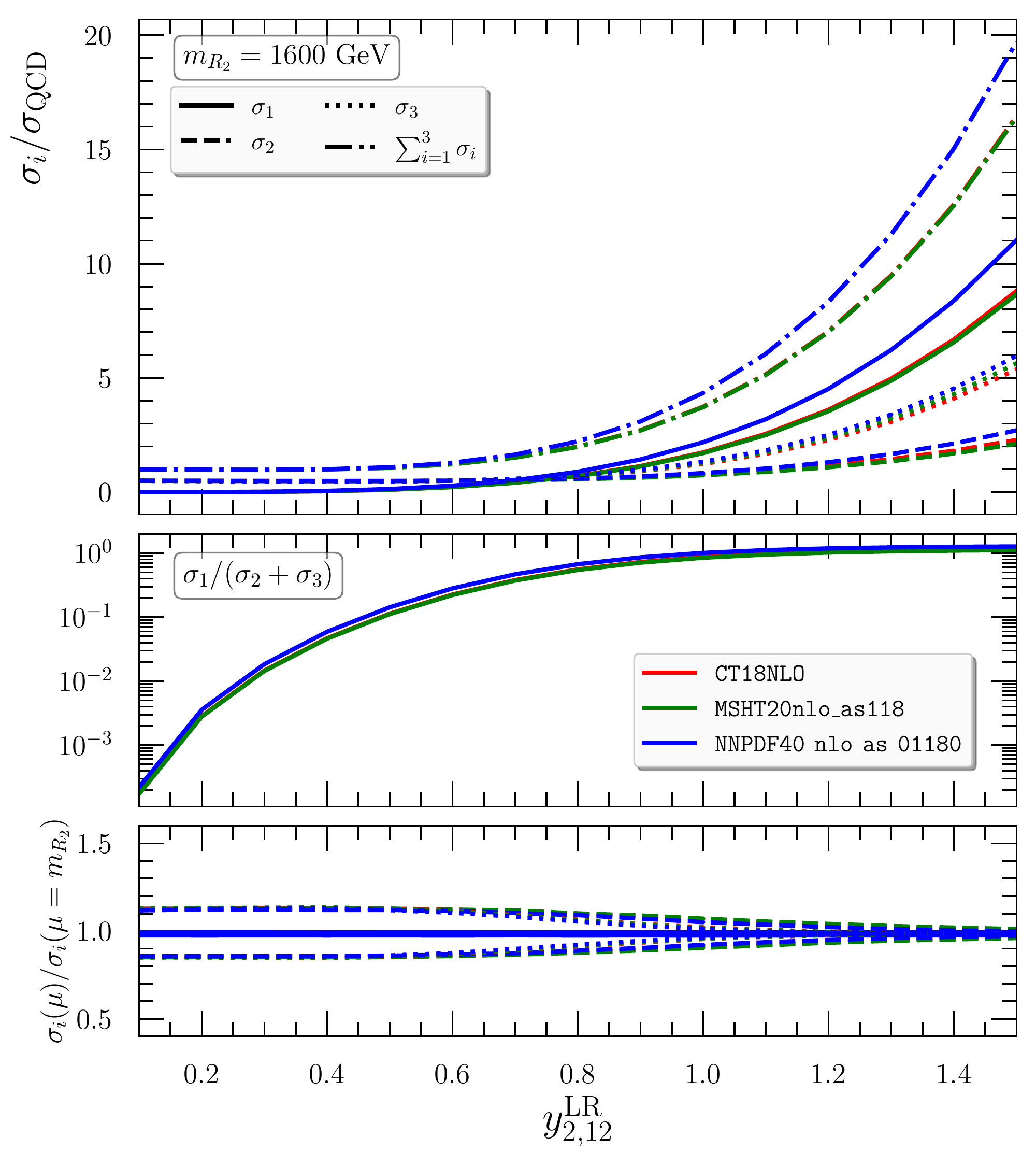}
    \caption{{\it Upper panels}: ratios of various $R_2$ leptoquark pair production cross sections of~\eqref{eq:sigma} to the QCD-driven cross section \eqref{eq:sigQCD}. Predictions are presented as a function of $y_{2,12}^{\rm LR}$, for $m_{\rm LQ} = 1600~{\rm GeV}$,  both at LO (left) and NLO (right). {\it Middle panels}: ratio of the off-diagonal leptoquark pair production cross section $\sigma_1$ to the total diagonal pair production cross section $\sigma_2 + \sigma_3$. {\it Lower panels}: dependence of $\sigma_i$ (for $i=1,2,3$) on the renormalisation and factorisation scales when they are varied simultaneously by a factor of 2 up and down.} 
    \label{fig:ratio:R2:y12:MLQ:1600} 
 \end{figure}   
 
In this section, we discuss the relevance of the off-diagonal leptoquark production modes with respect to the diagonal ones, for which we distinguish the pure QCD contributions from the `full' total cross section including both QCD and $t$-channel diagrams. We begin our study with the second $R_2$--$S_3$ scenario considered in section~\ref{sec:benchmark_generic}, in which all $S_3$ leptoquark eigenstates are decoupled. We correspondingly display results obtained in an `$R_2$-only' model, as a function of a single non-zero Yukawa coupling that is taken to be either $y_{2,12}^{\rm LR}$ or $y_{2,22}^{\rm LR}$. We then investigate the impact of both leptoquark species in the first scenario introduced in section~\ref{sec:benchmark_generic}, for which we show predictions depending on $y_{2,12}^{\rm LR} = y_{2,12}^{\rm RL} = y_{3,12}^{\rm LL}$.

In figure~\ref{fig:ratio:R2:y12:MLQ:1600} we estimate the impact of the $t$-channel contributions on the various leptoquark pair production channels (upper panel of the figures). We consider LO (left) and NLO (right) total rates associated with the two diagonal channels as well as with the single off-diagonal channel relevant for a model in which only an $R_2$ leptoquark is included,
\be\bsp
  \sigma_1\equiv&\ \sigma\Big[pp\to R_2^{(\pm5/3)} R_2^{(\mp2/3)}\Big],\\
  \sigma_2\equiv&\ \sigma\Big[pp\to R_2^{(-2/3)} R_2^{(+2/3)}\Big],\\
  \sigma_3\equiv&\ \sigma\Big[pp\to R_2^{(-5/3)} R_2^{(+5/3)}\Big]\,.
\esp\label{eq:sigma}\ee
The results are presented in the form of ratios to the cross section $\sigma_{\rm QCD}$ associated with the QCD-driven production of identical $R_2$ states,
\be 
 \sigma_{\rm QCD}
   \equiv \sigma\Big[pp\to R_2^{(-5/3)} R_2^{(+5/3)}\Big]
   + \sigma\Big[pp\to R_2^{(-2/3)} R_2^{(+2/3)}\Big]
 \quad \text{for}\quad y_{2,12}^{\rm LR}=0\,,
\label{eq:sigQCD}\ee
and we study the dependence of these ratios on the Yukawa coupling $y_{2,12}^{\rm LR}$ (all other Yukawa couplings being set to zero). In the middle panel of the figures, we investigate the importance of the off-diagonal cross section $\sigma_1$ relative to the sum of the diagonal ones $\sigma_2+\sigma_3$ (with $t$-channel contributions included), whereas the lower panel of the figures is dedicated to the scale uncertainties inherent to each of the three processes. Moreover, all results are shown for all PDF choices considered.

When $y_{2,12}^{\rm LR}$ is small, the ratios $\sigma_{2,3}/\sigma_{\rm QCD}$ are close to 0.5 for each of the diagonal modes, while $\sigma_1/\sigma_{\rm QCD}$ is close to 0 in the off-diagonal case. The $t$-channel effects are indeed small and negligible relative to the QCD rates, that only depend on the leptoquark mass. For larger $y_{2,12}^{\rm LR}$ values, $t$-channel contributions kick in and quickly dominate with increasing Yukawa couplings. Cross sections corresponding to the diagonal production rates become much larger than in the pure QCD case, and the off-diagonal mode even gets larger than the diagonal channels. In other words, the rise of the cross section due to $t$-channel contributions in a specific diagonal production mode is smaller than the rise of the cross section of the off-diagonal mode, whilst both these rises are significant. In addition, the middle panel of the figure indicates that the off-diagonal cross section is mostly equal to the sum of the cross sections of the two diagonal modes (including $t$-channel contributions) for large $y_{2,12}^{\rm LR}$ values.  

Comparing NLO (right panel) with LO (left panel) predictions, it turns out that LO-based analyses are not reliable to judge the importance of both the off-diagonal and the diagonal modes with respect to the pure QCD case (when the Yukawa couplings are moderate or large). NLO corrections are indeed very relevant for all modes, and they affect differently QCD-driven and $t$-channel-driven diagrams. The $\sigma_1/(\sigma_2+\sigma_3)$ ratio stays however relatively insensitive to the precision of the pertubative calculation (middle panel of the figures). In addition, this ratio is quite robust relative to the chosen set of PDFs, and this is even more true at NLO. On the other hand, the ratios in the upper panels of the plots significantly disagree with each other when the employed PDF set is varied. 

In the lower panels of the figures, we show the relative size of the theoretical errors originating from scale variation for each of the three production modes, and for the different PDF set considered. Receiving a significant contribution from pure QCD diagrams, the diagonal modes are in principle plagued with a bigger theoretical error than the off-diagonal one. However, at large values of $y_{2,12}^{\rm LR}$, \ie~when the importance of the pure QCD contributions diminishes, the size of the theoretical error gradually becomes independent of the production channel. Finally, as expected the size of scale variation errors gets significantly smaller at NLO (by at least a factor of 3), and is mostly independent of the chosen PDF set.  

 \begin{figure}
    \includegraphics[width=0.48\linewidth]{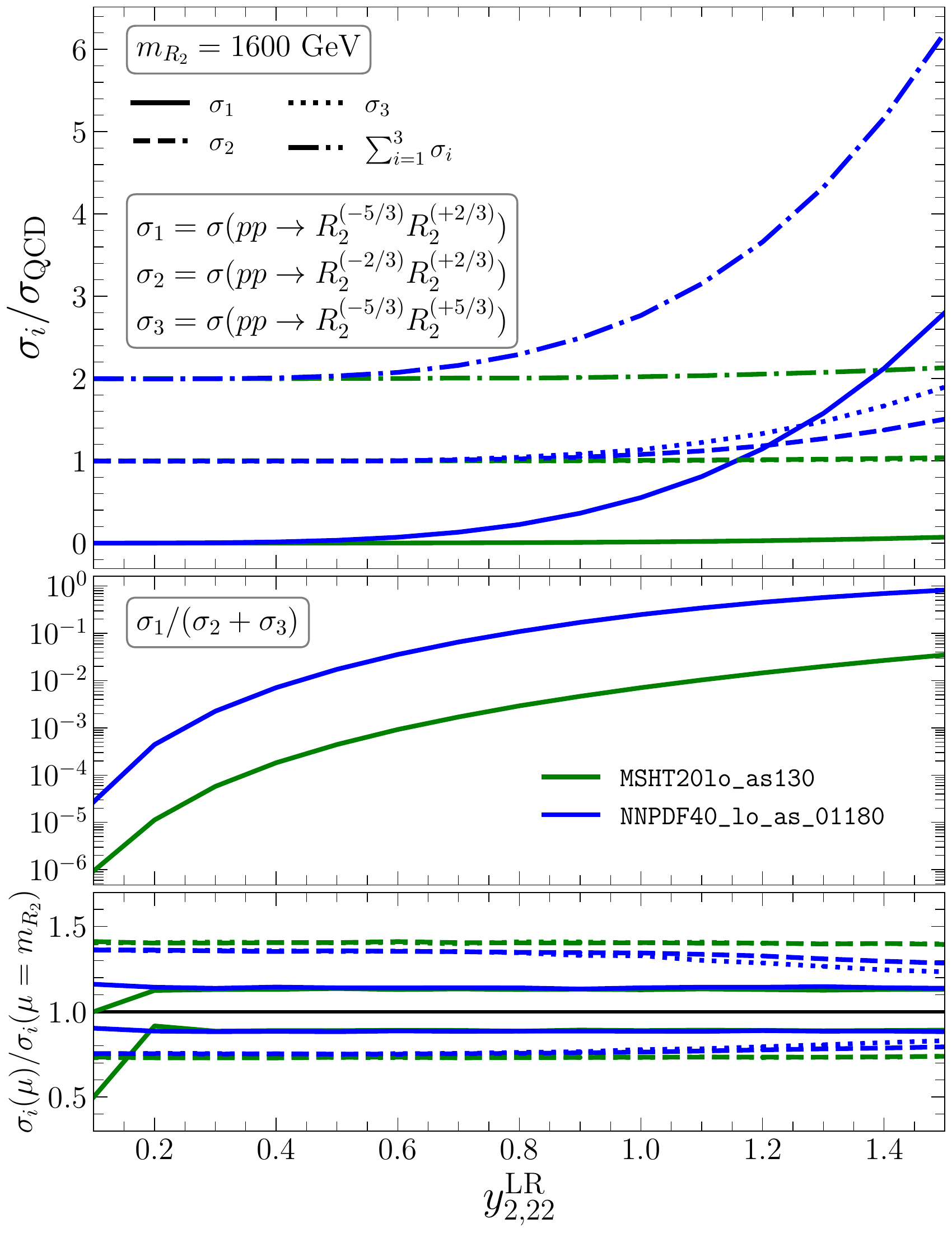}
    \hfill
    \includegraphics[width=0.48\linewidth]{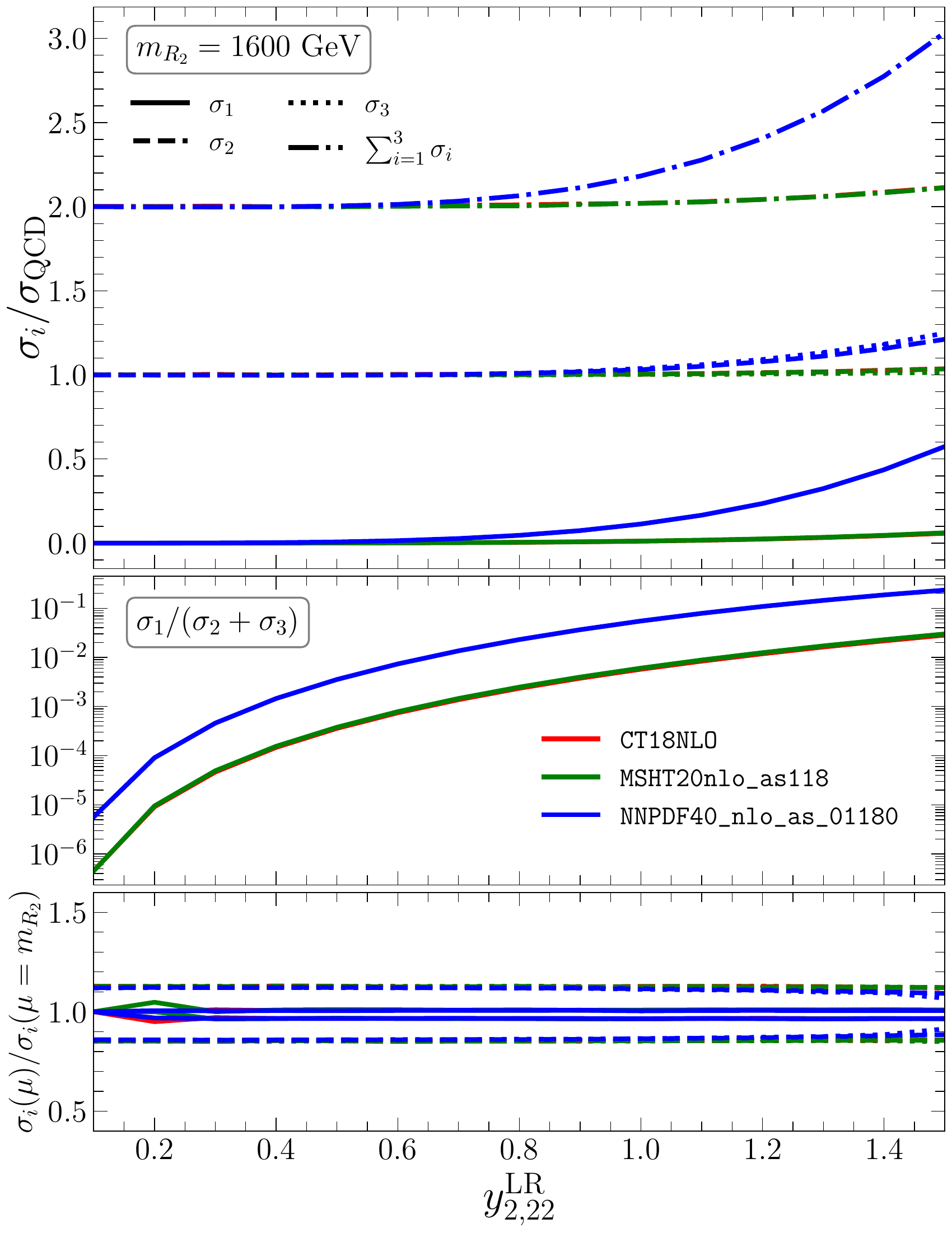}
    \caption{Same as in figure \ref{fig:ratio:R2:y12:MLQ:1600} but for an $R_2$ leptoquarks scenario in which the only non-zero Yukawa coupling is $y_{2,22}^{\rm LR}$.}
    \label{fig:ratio:R2:y22:MLQ:1600}
\end{figure}

In figure~\ref{fig:ratio:R2:y22:MLQ:1600}, we display the results of a similar study, but for an $R_2$ leptoquark scenario in which $y_{2,22}^{\rm LR}$ is left free to be varied and all other Yukawa couplings are fixed to zero. The difference in the behaviour of the various cross sections from eq.~\eqref{eq:sigma} relative to the results shown in figure~\ref{fig:ratio:R2:y12:MLQ:1600} is quite dramatic, and it is mostly due to the different possibilities for the initial states. At tree-level and when solely the Yukawa coupling $y_{2,22}^{\rm LR}$ is non-zero, there is no option for a $t$-channel subprocess involving an initial valence quark to contribute. This reduces the potential impact of all (diagonal and off-diagonal) $t$-channel contributions, as visible from the upper panels of the figure. The ratios of the three leptoquark production rates to the pure QCD result~\eqref{eq:sigQCD} are indeed much lower at large values of the Yukawa coupling than in figure~\ref{fig:ratio:R2:y12:MLQ:1600}.  In addition, the diagonal production modes dominate, with cross sections that are a few times larger than that of the off-diagonal mode (see the middle panels of the figure). Accordingly, theoretical uncertainties due to scale variation are almost independent of $y_{2,22}^{\rm LR}$ for the diagonal channels, and larger in these cases than in the off-diagonal case.

Our results also depict a much higher dependence on the PDF set used in the calculations. Predictions obtained with NNPDF4.0 densities are indeed much different from those obtained with either CT18 or MSHT20 PDFs (that agree with each other at NLO). For instance, the ratio $\sigma_1/(\sigma_2+\sigma_3)$ differs at NLO by as much as one order of magnitude when comparing MSHT20 and NNPDF4.0 predictions. We attribute these large difference to the treatment of the charm quark distribution in the PDF fitting procedure, that is independently parametrised in the default NNPDF4.0 case and fully perturbatively treated in the CT18 and MSHT20 cases (see also section~\ref{sec:BSs}).

\begin{figure}
    \centering
    \includegraphics[width=0.49\linewidth, height=12cm]{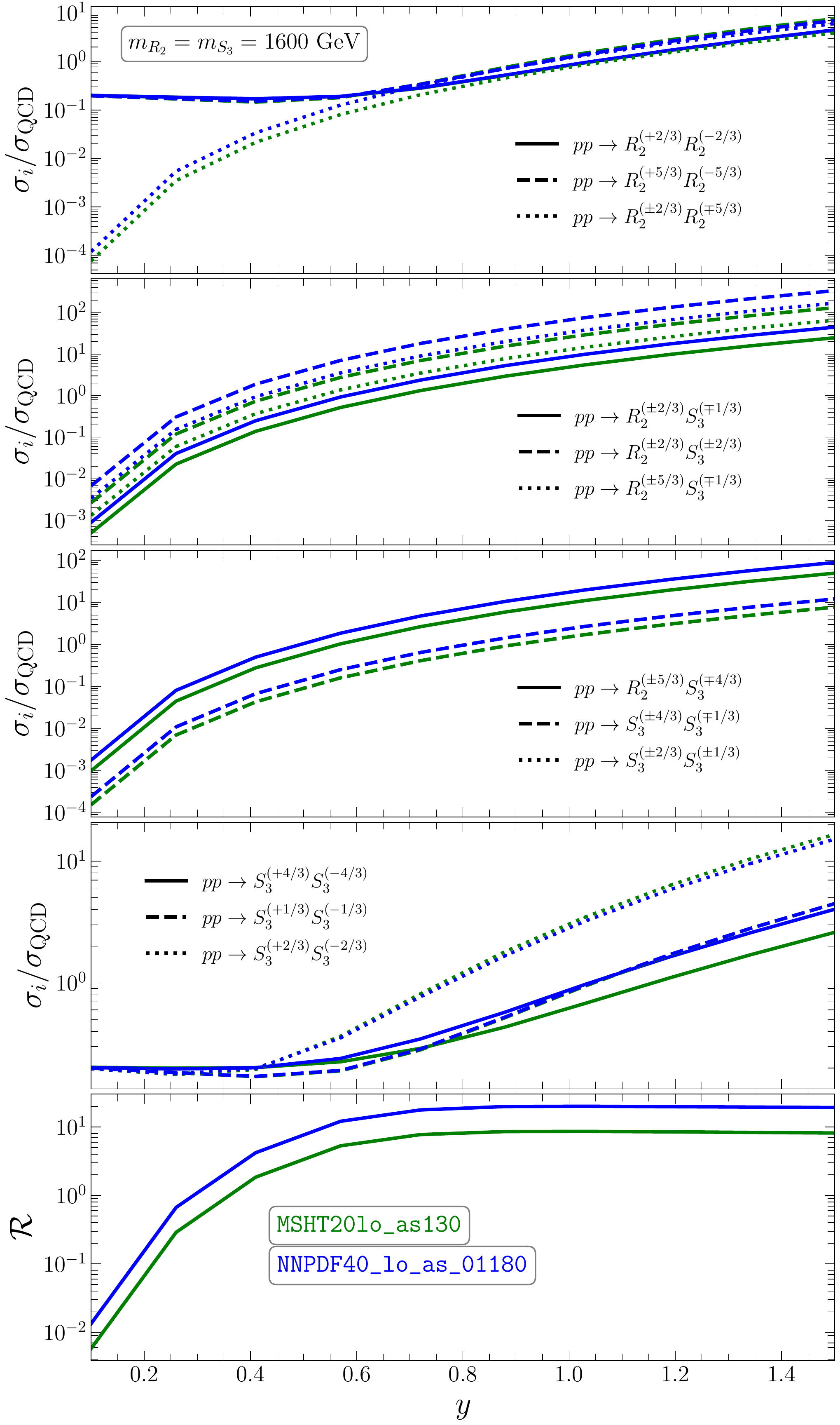}
    \hfill
    \includegraphics[width=0.49\linewidth, height=12cm]{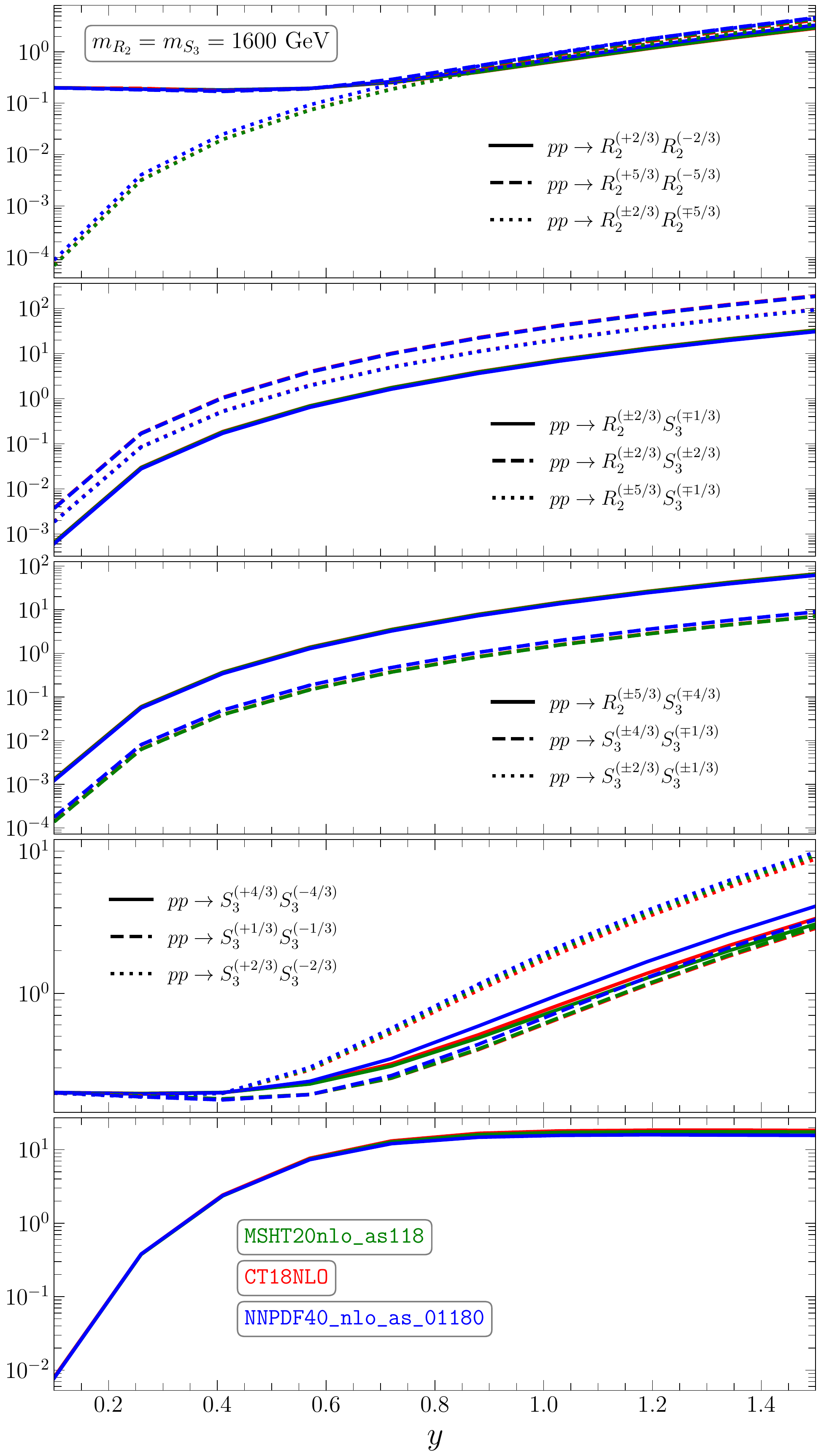}
    \caption{{\it Upper four panels}: ratios of the different scalar leptoquark pair production cross section with respect to the pure QCD contribution as a function of $y = y_{2,12}^{\rm LR} = y_{2,12}^{\rm RL} = y_{3, 12}^{\rm LL}$ at LO (left) and NLO (right). $\sigma_{\rm QCD}$ is defined as the sum of the cross sections for all diagonal channels with $y = 0$, all predictions correspond to $m_{R_2} = m_{S_3} = 1600~{\rm GeV}$. {\it Lower panel}: ratio~\eqref{eq:ratioR} of the (full) off-diagonal production rate to (full) the diagonal one as a function of $y$.}
    \label{fig:ratios:R2S3:mLQ:1600}
\end{figure}

After studying the relevance of the off-diagonal modes in the simpler $R_2$ model with only one non-zero Yukawa coupling, we consider cross sections in the $R_2$--$S_3$ model in figure~\ref{fig:ratios:R2S3:mLQ:1600}. In the top four panels of the figure we present ratios of the production cross sections associated with a specific process to the sum all pure QCD contributions, at LO (left figure) and NLO (right figure). Here,
\begin{eqnarray}
 \sigma_{\rm QCD} &\equiv& 
     \sigma\Big[pp\to R_2^{(-5/3)} R_2^{(+5/3)}\Big]
   + \sigma\Big[pp\to R_2^{(-2/3)} R_2^{(+2/3)}\Big]
   + \sigma\Big[pp\to S_3^{(-4/3)} S_3^{(+4/3)}\Big] \nonumber \\
   && {}+ \sigma\Big[pp\to S_3^{(-2/3)} S_3^{(+2/3)}\Big]
   + \sigma\Big[pp\to S_3^{(-1/3)} S_3^{(+1/3)}\Big]  \quad \text{for}\quad y = 0.
\end{eqnarray}
NLO corrections can lead to a significant change of the ratios, often by tens of percent or more. Furthermore, at high enough values of the Yukawa coupling off-diagonal channels and and $t$-channel diagonal contributions dominate over pure QCD production. This dominance reaches factors of 10--100, and is most pronounced for off-diagonal processes involving two valence quarks. 

The relative importance of the off-diagonal modes can be also judged by considering the ratio of the cross section for all possible off-diagonal production modes allowed in the model to that of all diagonal production modes (including $t$-channel contributions). This ratio is defined by
\begin{eqnarray}\label{eq:ratioR}
{\cal R} = \frac{\sum_{i\neq j} \sigma\left[pp \to {\rm LQ}_i\, {\rm LQ}_j\right]}{\sum_i \sigma\left[pp \to {\rm LQ}_i\, {\rm LQ}_i\right]}.
\end{eqnarray}
and is shown in the lowest panel in figure~\ref{fig:ratios:R2S3:mLQ:1600}. In the particular model considered here,  the off-diagonal-channels become as important as the diagonal ones at an already relative small values of $y\sim 0.35$. Furthermore, at $y\gtrsim 1$ off-diagonal leptoquark pair production is around 20 times higher than diagonal leptoquark pair production.  Comparing with the previously studied $R_2$ model, we observe that this behaviour is entirely driven by the possibility to produce a pair of leptoquarks from an initial state including one or more valence quarks. Whereas in the  $R_2$ model the off-diagonal  $R_2^{(\pm2/3)}R_2^{(\mp5/3)}$ production involves at LO a $d \bar u/\bar du$ initial state, $R_2^{(\pm2/3)}S_3^{(\pm2/3)}$ production originates from $uu/\bar u \bar u$ scattering.

Results for the same scenarios as discussed in this section, but with a higher leptoquark mass of $m_{\rm LQ}=2400$ GeV, are presented in appendix~\ref{app:highermass}.

%%%%%%%%%%%%%%%%%%%%%%%%%
\section{Role of the off-diagonal production modes in scenarios addressing the flavour anomalies}
\label{sec:BSs}
%%%%%%%%%%%%%%%%%%%%%%%%%
In this section, we discuss predictions for scalar leptoquark production processes in the context of the benchmark scenarios introduced in section~\ref{sec:benchmarksFlavor}, and we evaluate the impact of the off-diagonal production channels relatively to the diagonal ones. Whereas the latter have been already studied in \cite{Borschensky:2021hbo}, we provide predictions for completeness. Moreover, in the case of results obtained with the NNPDF set, we update our predictions by making use of the latest version 4.0 of their fit.

%%%%%%%%%%%%%%%%%%%%%%%%%%%%%%%%%%%
\subsection{Total cross sections at NLO}
%%%%%%%%%%%%%%%%%%%%%%%%%%%%%%%%%%%

\begin{table}
\renewcommand{\arraystretch}{1.2}
\setlength\tabcolsep{5pt}
\centering
\begin{tabular}{llccc}
& & \texttt{CT18NLO} & \texttt{MSHT20\_nlo\_as\_0118} & \texttt{NNPDF40\_nlo\_as\_01180}  \\
\midrule
\multirow{2}{*}{$a_1$} &
 $R_2^{(+2/3)} R_2^{(-2/3)}$ & $6.89 \substack{+7.7\% \\ -9.6\%} \substack{+18.8\% \\ -13.7\%}$ & $6.82 \substack{+7.2\% \\  -10.6\%} \substack{+8.2\% \\ -5.7\%}$ & $11.17 \substack{+3.0\% \\ -4.2\%} \substack{+25.2\% \\ -25.2\%}$ \\
& $R_2^{(+5/3)} R_2^{(-5/3)}$ & $6.87 \substack{+8.3\% \\ -9.8\%} \substack{+18.9\% \\ -13.7\%}$ & $6.86 \substack{+7.9\% \\ -10.2\%} \substack{+8.2\% \\ -5.7\%}$ & $11.15 \substack{+3.5\% \\ -4.0\%} \substack{+25.2\% \\ -25.2\%}$ \\
\midrule
\multirow{2}{*}{$a_2$} & $R_2^{(+2/3)} R_2^{(-2/3)}$ & $5.49 \substack{+12.2\% \\  -13.0\%} \substack{+13.8\% \\ -10.4\%}$ & $5.56 \substack{+13.2\% \\  -11.7\%} \substack{+5.9\% \\ -4.1\%}$ & $5.37 \substack{+10.7\% \\ -12.4\%} \substack{+2.7\% \\ -2.7\%}$\\
& $R_2^{(+5/3)} R_2^{(-5/3)}$ & $5.48 \substack{+11.2\% \\  -13.1\%} \substack{+13.8\% \\ -10.4\%}$ & $5.57 \substack{+13.5\% \\  -11.7\%} \substack{+5.9\% \\ -4.1\%}$ & $5.37 \substack{+10.4\% \\ -12.9\%} \substack{+2.7\% \\ -2.7\%}$ \\
\end{tabular}
\caption{Total cross sections at NLO-QCD (in fb) for the $a_1$ and $a_2$ benchmark points of table~\ref{tab:benchmarksR2}. Our predictions include  $t$-channel contributions, scale uncertainties (first errors reported) and PDF uncertainties (second errors reported). Cross sections below 1~ab are not shown.}\vspace{.4cm}
\label{tab:xsec:a1a2}
\begin{tabular}{llccc}
& & \texttt{CT18NLO} & \texttt{MSHT20\_nlo\_as\_0118} & \texttt{NNPDF40\_nlo\_as\_01180}  \\
\midrule
\multirow{7}{*}{$b_1$} & $R_2^{(+2/3)} R_2^{(-2/3)}$ & $0.708 \substack{+11.3\% \\ -14.3\%} \substack{+18.1\% \\ -13.4\%}$ & $0.712 \substack{+12.7\% \\ -12.9\%} \substack{+7.9\% \\ -5.4\%}$ & $0.714 \substack{+10.0\% \\  -12.2\%} \substack{+4.9\% \\ -4.9\%}$ \\
& $R_2^{(+5/3)} R_2^{(-5/3)}$ & $0.706 \substack{+11.1\% \\ -13.9\%} \substack{+18.1\% \\ -13.4\%}$ & $0.712
\substack{+13.3\% \\ -13.3\%} \substack{+7.9\% \\ -5.4\%}$ & $0.715 \substack{+9.5\% \\  -12.8\%} \substack{+4.8\% \\ -4.8\%}$ \\
& $S_3^{(+2/3)} S_3^{(-2/3)}$ & $0.0107 \substack{+15.4\% \\ -14.9\%} \substack{+30.5\% \\ -21.7\%}$ & $0.0109 \substack{+15.9\% \\ -14.3\%} \substack{+14.5\% \\ -9.8\%}$ & $0.0110 \substack{+12.3\% \\ -14.7\%} \substack{+9.1\% \\ -9.1\%}$ \\
& $S_3^{(+1/3)} S_3^{(-1/3)}$ & $0.0108 \substack{+14.6\% \\ -15.7\%} \substack{+30.5\% \\ -21.7\%}$ & $0.0110 \substack{+15.7\% \\ -14.5\%} \substack{+14.5\% \\ -9.8\%}$ & $0.0111 \substack{+12.0\% \\ -14.9\%} \substack{+9.1\% \\ -9.1\%}$ \\
& $S_3^{(+4/3)} S_3^{(-4/3)}$ & $0.0108 \substack{+14.2\% \\ -16.0\%} \substack{+30.7\% \\ -21.8\%}$ & $0.0110 \substack{+15.2\% \\ -14.8\%} \substack{+14.7\% \\ -9.9\%}$ & $0.0111 \substack{+11.6\% \\ -14.5\%} \substack{+9.1\% \\ -9.1\%}$ \\
& $R_2^{(\pm5/3)} S_3^{(\mp4/3)}$ & $0.0014 \substack{+5.0\% \\ -5.2\%} \substack{+72.1\% \\ -46.2\%}$ & $0.00144 \substack{+8.9\% \\ -2.2\%} \substack{+32.2\% \\ -21.3\%}$ & $0.0066 \substack{+4.4\% \\ -5.5\%} \substack{+44.3\% \\ -44.3\%}$ \\
& $R_2^{(\pm2/3)} S_3^{(\mp1/3)}$ & $0.00069 \substack{+5.0\% \\ -5.2\%} \substack{+72.1\% \\ -46.1\%}$ & $0.00072 \substack{+8.6\% \\ -2.5\%} \substack{+32.1\% \\ -21.2\%}$ & $0.0033 \substack{+4.7\% \\ -5.3\%} \substack{+44.2\% \\ -44.2\%}$ \\
\midrule
\multirow{7}{*}{$b_2$} & $R_2^{(+2/3)} R_2^{(-2/3)}$ & $0.719 \substack{+11.8\% \\ -13.4\%} \substack{+18.9\% \\ -13.9\%}$ & $0.726  \substack{+13.2\% \\ -12.1\%} \substack{+8.3\% \\ -5.7\%}$ & $0.832 \substack{+8.1\% \\ -10.8\%} \substack{+11.9\% \\ -11.9\%}$  \\
& $R_2^{(+5/3)} R_2^{(-5/3)}$ & $0.719 \substack{+11.5\% \\ -13.3\%} \substack{+18.8\% \\ -13.8\%}$ & $0.727 \substack{+12.9\% \\ -12.6\%} \substack{+8.2\% \\ -5.6\%}$ & $0.834 \substack{+7.5\% \\ -10.8\%} \substack{+11.9\% \\ 11.9\%}$   \\
& $S_3^{(+2/3)} S_3^{(-2/3)}$ & $0.0107 \substack{+15.4\% \\ -15.0\%} \substack{+30.5\% \\ -21.7\%}$ & $0.0109 \substack{+15.9\% \\ -14.3\%} \substack{+14.5\% \\ -9.8\%}$ & $0.0110 \substack{+12.7\% \\ -14.7\%} \substack{+9.1\% \\ -9.1\%}$ \\
& $S_3^{(+1/3)} S_3^{(-1/3)}$ & $0.0108 \substack{+14.3\% \\ -15.8\%} \substack{+30.7\% \\ -21.8\%}$ & $0.0110 \substack{+15.6\% \\ -14.3\%} \substack{+14.6\% \\  -9.9\%}$ & $0.0108 \substack{+11.5\% \\ -14.7\%} \substack{+9.1\% \\ -9.1\%}$ \\
& $S_3^{(+4/3)} S_3^{(-4/3)}$ & $0.0109 \substack{+14.0\% \\ -15.1\%} \substack{+31.5\% \\ -22.2\%}$ & $0.0111 \substack{+15.6\% \\ -14.1\%}   \substack{+15.1\% \\ -10.2\%}$ & $0.0115 \substack{+11.8\% \\ -14.2\%} \substack{+9.1\% \\ -9.1\%}$ \\
& $R_2^{(\pm5/3)} S_3^{(\mp4/3)}$ & $0.0048 \substack{+4.9\% \\ -5.4\%} \substack{+72.2\% \\ -46.2\%}$ & $0.0049 \substack{+9.3\% \\ -2.4\%} \substack{+32.1\% \\  -21.3\%}$ & $0.0227 \substack{+4.4\% \\ -5.6\%} \substack{+44.2\% \\ -44.2\%}$ \\
& $R_2^{(\pm2/3)} S_3^{(\mp1/3)}$ & $0.0024 \substack{+4.9\% \\ -5.2\%} \substack{+72.1\% \\ -46.1\%}$ &  $0.0025 \substack{+8.4\% \\ -2.5\%}  \substack{+32.1\% \\ -21.3\%}$ & $0.0113 \substack{+4.6\% \\ -5.3\%} \substack{+44.2\% \\ -44.2\%}$ \\
\end{tabular}
\caption{Same as table~\ref{tab:xsec:a1a2} but for the $b_1$ and $b_2$ scenarios of table~\ref{tab:benchmarksR2plusS3}.}
\label{tab:xsec:b1b2}
\end{table}

\begin{table}
\setlength\tabcolsep{5pt}
\centering
\begin{tabular}{llccc}
& & \texttt{CT18NLO} & \texttt{MSHT20\_nlo\_as\_0118} & \texttt{NNPDF40\_nlo\_as\_01180}  \\
\midrule
\multirow{8}{*}{$c_1$} & $S_1^{(+1/3)} S_1^{(-1/3)}$ & $1.45 \substack{+10.4\% \\ -12.8\%} \substack{+18.7\% \\ -13.0\%}$ & $1.47 \substack{+12.3\% \\ -10.9\%} \substack{+7.8\% \\ -5.4\%}$ & $1.78 \substack{+6.6\% \\ -8.9\%} \substack{+10.8\% \\ -10.8\%}$ \\
& $S_3^{(+2/3)} S_3^{(-2/3)}$ & $1.35 \substack{+11.3\% \\ -14.3\%}  \substack{+16.3\% \\ -12.2\%}$ & $1.36 \substack{+12.9\% \\ -12.8\%} \substack{+7.1\% \\ -4.9\%}$ & $1.30 \substack{+10.9\% \\ -13.2\%} \substack{+3.5\% \\ -3.5\%}$ \\
& $S_3^{(+1/3)} S_3^{(-1/3)}$ & $1.35 \substack{+11.9\% \\ -13.9\%} \substack{+16.3\% \\ -12.2\%}$ & $1.36 \substack{+13.4\% \\ -13.0\%} \substack{+7.1\% \\ -4.9\%}$ & $1.29 \substack{+10.7\% \\ -14.3\%} \substack{+3.5\% \\ -3.5\%}$ \\
& $S_3^{(+4/3)} S_3^{(-4/3)}$ & $1.35 \substack{+11.9\% \\ -13.6\%} \substack{+16.4\% \\ -12.2\%}$ & $1.37  \substack{+12.6\% \\ -12.8\%} \substack{+7.1\% \\ -4.9\%}$ & $1.30 \substack{+10.8\% \\ -13.1\%} \substack{+3.5\% \\ -3.5\%}$ \\
& $S_1^{(\pm1/3)} S_3^{(\mp1/3)}$ & $0.036 \substack{+4.3\% \\ -5.4\%} \substack{+115.0\% \\ -33.1\%}$ & $0.038 \substack{+10.2\% \\ -0.1\%} \substack{+24.5\% \\ -19.6\%}$ & $0.14 \substack{+4.1\% \\ -5.8\%} \substack{+35.7\% \\ -35.7\%}$ \\
& $S_1^{(\pm1/3)} S_3^{(\pm2/3)}$ & $0.032 \substack{+4.4\% \\ -5.2\%} \substack{+71.7\% \\ -30.8\%}$ & $0.034 \substack{+9.9\% \\ -0.2\%} \substack{+22.7\% \\ -15.8\%}$ & $0.15 \substack{+4.6\% \\ -5.2\%} \substack{+38.2\% \\ -38.2\%}$ \\
& $S_3^{(\pm2/3)} S_3^{(\pm1/3)}$ & $0.0051 \substack{+4.9\% \\ -4.4\%} \substack{+71.3\% \\ -30.8\%}$ & $0.0054 \substack{+10.7\% \\ -0.1\%} \substack{+22.7\% \\ -15.8\%}$ & $0.0239 \substack{+4.3\% \\ -5.7\%} \substack{+38.2\% \\ -38.2\%}$ \\
& $S_3^{(\pm1/3)} S_3^{(\mp4/3)}$ & $0.0051 \substack{+4.9\% \\ -4.4\%} \substack{+71.1\% \\ -30.8\%}$ & $0.0054 \substack{+10.8\% \\ -0.2\%} \substack{+22.8\% \\ -15.7\%}$ & $0.0239 \substack{+4.5\% \\ -5.8\%} \substack{+38.1\% \\ -38.1\%}$\\
\midrule
\multirow{8}{*}{$c_2$} & $S_1^{(+1/3)} S_1^{(-1/3)}$ & $1.44 \substack{+10.4\% \\ -12.9\%} \substack{+18.2\% \\ -13.0\%}$ & $1.46 \substack{+12.0\% \\  -11.8\%} \substack{+7.7\% \\ -5.3\%}$ &  $1.73 \substack{+7.2\% \\ -9.1\%} \substack{+10.8\% \\ -10.8\%}$ \\
& $S_3^{(+2/3)} S_3^{(-2/3)}$ & $1.41 \substack{+10.6\% \\ -13.4\%} \substack{+17.5\% \\ -13.0\%}$ & $1.41 \substack{+12.1\% \\  -12.1\%} \substack{+7.6\% \\ -5.3\%}$ &  $1.64 \substack{+6.9\% \\  -10.1\%} \substack{+12.1\% \\ -12.1\%}$ \\
& $S_3^{(+1/3)} S_3^{(-1/3)}$ & $1.39 \substack{+11.1\% \\ -13.3\%} \substack{+17.0\% \\ -12.5\%}$ & $1.40 \substack{+12.9\% \\  -11.9\%} \substack{+7.3\% \\ -5.0\%}$ & $1.47 \substack{+9.4\% \\ -11.9\%} \substack{+5.1\% \\ -5.1\%}$  \\
& $S_3^{(+4/3)} S_3^{(-4/3)}$ & $1.48 \substack{+10.9\% \\ -12.2\%} \substack{+22.1\% \\ -13.0\%}$ & $1.52 \substack{+13.0\% \\  -9.7\%} \substack{+7.9\% \\ -5.5\%}$ & $1.74 \substack{+7.2\% \\ -9.4\%} \substack{+10.0\% \\ -10.0\%}$ \\
& $S_1^{(\pm1/3)} S_3^{(\mp1/3)}$ & $0.138 \substack{+4.7\% \\ -5.1\%} \substack{+115.0\% \\ -33.1\%}$ & $0.146 \substack{+9.7\% \\  -0.2\%} \substack{+24.6\% \\ -19.7\%}$ & $0.553 \substack{+4.1\% \\ -5.8\%} \substack{+35.8\% \\ -35.8\%}$ \\
& $S_1^{(\pm1/3)} S_3^{(\pm2/3)}$ & $0.125 \substack{+4.4\% \\ -6.7\%} \substack{+71.6\% \\ -30.9\%}$ & $0.132 \substack{+9.8\% \\  -0.1\%} \substack{+22.7\% \\ -15.8\%}$ & $0.588 \substack{+4.6\% \\  -5.3\%} \substack{+38.2\% \\ -38.2\%}$ \\
& $S_3^{(\pm2/3)} S_3^{(\pm1/3)}$ & $0.102 \substack{+5.0\% \\ -4.8\%} \substack{+71.9\% \\ -30.8\%}$ & $0.108 \substack{+11.0\% \\  -0.2\%} \substack{+23.0\% \\ -15.8\%}$ & $0.481 \substack{+4.3\% \\  -5.7\%} \substack{+38.1\% \\ -38.1\%}$ \\
& $S_3^{(\pm1/3)} S_3^{(\mp4/3)}$ & $0.102 \substack{+5.1\% \\ -4.2\%} \substack{+71.7\% \\ -30.9\%}$ & $0.108 \substack{+10.9\% \\  -0.5\%} \substack{+22.8\% \\ -15.8\%}$ & $0.481 \substack{+4.5\% \\  -5.8\%} \substack{+38.1\% \\ -38.1\%}$ \\
\end{tabular}
\caption{Same as table~\ref{tab:xsec:a1a2} but for the $c_1$ and $c_2$ scenarios of table~\ref{tab:benchmarksS1plusS3}.}
\label{tab:xsec:c1c2}
\end{table}

In tables~\ref{tab:xsec:a1a2}, \ref{tab:xsec:b1b2}, and \ref{tab:xsec:c1c2}, we show predictions for NLO-QCD total cross sections of all leptoquark pair production modes relevant for the $R_2$, $R_2$--$S_3$, and $S_1$--$S_3$ scenarios of section~\ref{sec:benchmr2}, respectively. Our results include scale and PDF uncertainties, and we only consider processes for which the cross section is larger than 1~ab for at least one of the three PDF sets used.

Off-diagonal contributions are generally small for all selected benchmark points. For scenarios $a_1$ and $a_2$ in the $R_2$ model, they are indeed negligible, with rates falling below 1~ab. They are thus not shown in table~\ref{tab:xsec:a1a2}. For benchmarks $b_1$ and $b_2$ in the $R_2$--$S_3$ model, there exist two off-diagonal channels with cross sections larger than 1~ab, $pp\to R_2^{(\pm5/3)} S_3^{(\mp4/3)}$ and $pp\to R_2^{(\pm2/3)} S_3^{(\mp1/3)}$, for which the rates are therefore reported in table~\ref{tab:xsec:b1b2}. They consist of about 0.1\% to 2\% of the {\it full} leptoquark pair-production cross section (where all channels are summed over), the precise number depending on the benchmark point and on the employed PDF set. The scaling relation between the two off-diagonal channels is not surprising, and could be actually expected from eq.~\eqref{r2s3scaling}. For benchmarks $c_1$ and $c_2$ in the $S_1$--$S_3$ model, four channels are associated with rates larger than 1~ab, and are thus displayed in table~\ref{tab:xsec:c1c2}: $pp\to S_1^{(\pm1/3)} S_3^{(\mp1/3)}$, $pp\to S_1^{(\pm1/3)} S_3^{(\pm2/3)}$, $pp\to S_3^{(\pm2/3)} S_3^{(\pm1/3)}$, and $pp\to S_3^{(\pm1/3)} S_3^{(\mp4/3)}$. When CT18 and MSHT20 parton densities are used, they impact the {\it full} leptoquark pair production rate by 1\% to 8\%, depending on the scenario, and those values increase to 5\% and 25\% when the NNPDF4.0 set is employed. This large difference in the predictions can be attributed to the different treatment of the charm distribution in the NNPDF set, as discussed more extensively in section~\ref{sec:charm}. Moreover, the relative impact of the off-diagonal contributions is directly connected to the strength of the leptoquark coupling to second-generation quarks with respect to all other Yukawa couplings. Consequently, off-diagonal channels play a bigger role in scenarios $b_2$ and $c_2$ relatively to scenarios $b_1$ and $c_1$, due to the larger values of the couplings $y_{2,23}^{\rm RL}$ (in $b_2$) and $y_{3,23}^{\rm LL}$ (in $c_2$), respectively.

NLO corrections are also known to generally reduce scale uncertainties. In this context, it can  be seen that for all off-diagonal channels, scale uncertainties get smaller than these inherent to the diagonal channels. This originates from the absence of $\mathcal{O}(\alphas^2)$ contributions at LO in the off-diagonal case. On the contrary, PDF uncertainties in the off-diagonal channels are significantly increased by a factor of at least 2, due to the strong dependence of the rates on the heavy sea-quark distributions.

Our results therefore show that once NLO-QCD predictions are considered, off-diagonal leptoquark pair-production channels generally lead to an irrelevant contribution to the {\it full} leptoquark pair production rate for scenarios motivated by the flavour anomalies. Exceptions however exist, as shown with the benchmarks $c_1$ and $c_2$, where effects reaching tens of percent are observed.

%%%%%%%%%%%%%%%%%%%%%%%%%%%%%%%%%%%%%%%%%%%
\subsection{Impact of the charm PDF on the predictions}
\label{sec:charm}
%%%%%%%%%%%%%%%%%%%%%%%%%%%%%%%%%%%%%%%%%%%
\begin{figure}
    \centering
    \includegraphics[width=0.495\linewidth]{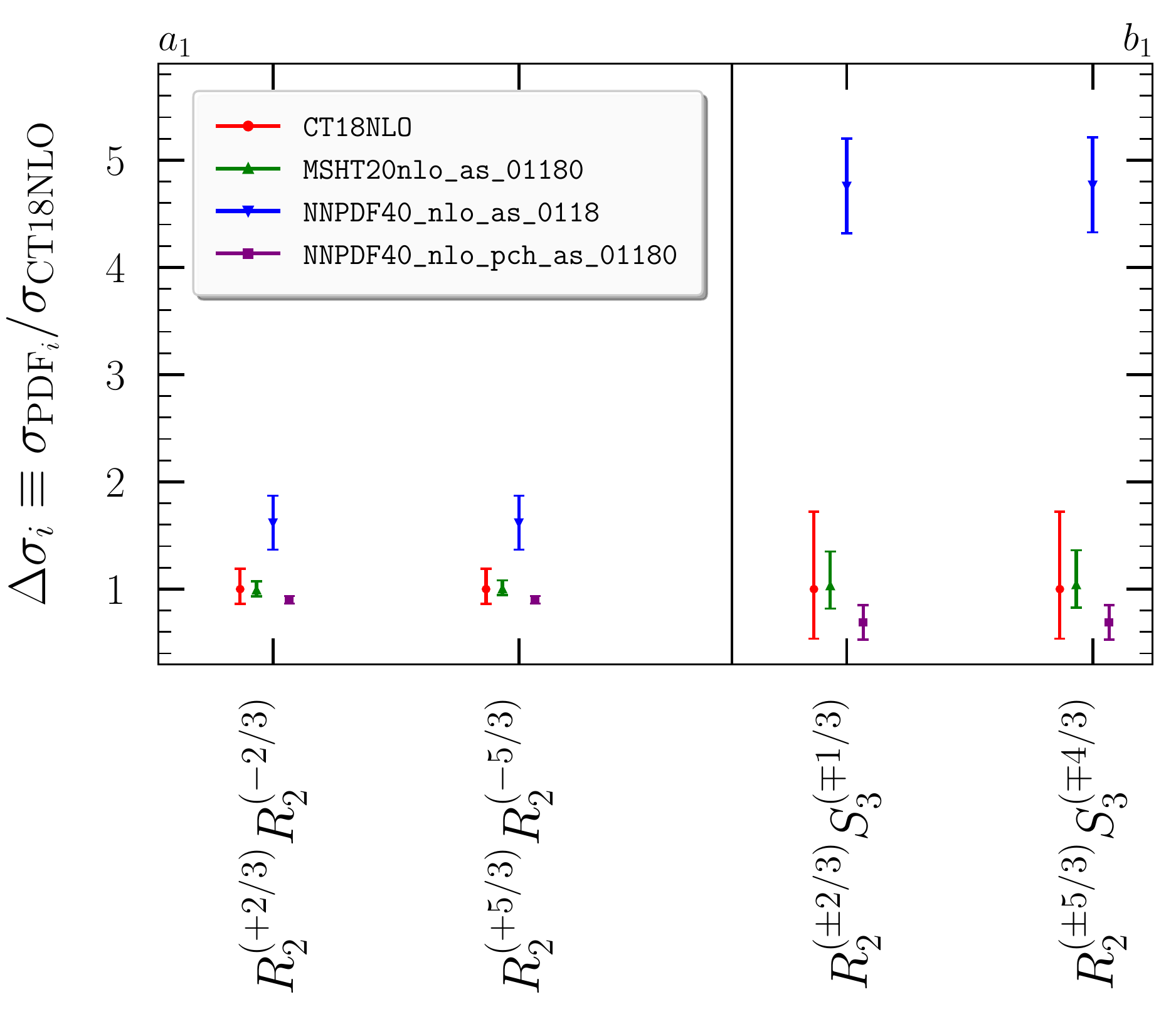}\hfill\includegraphics[width=0.495\linewidth]{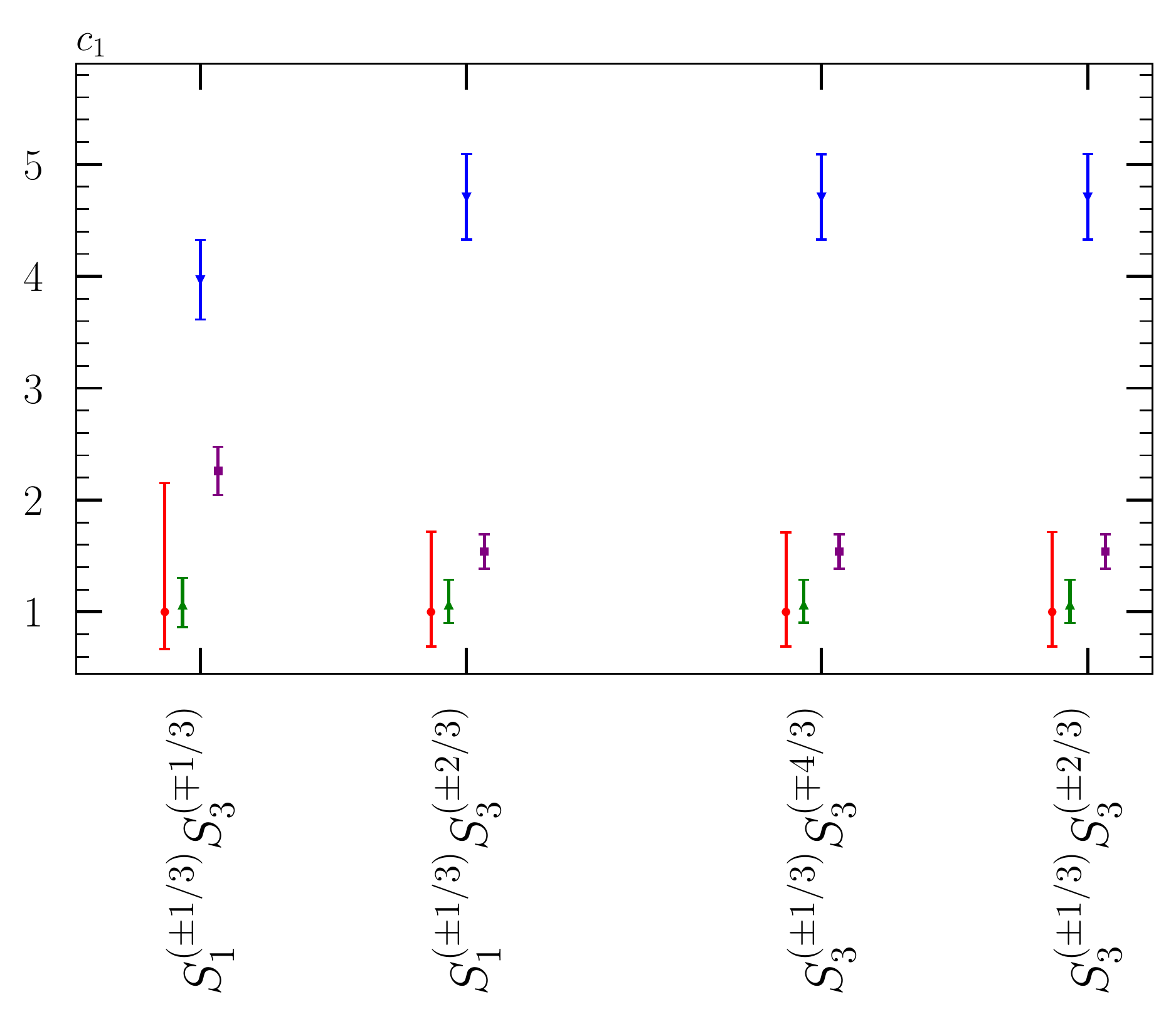}
    \caption{Total cross section predictions, with the associated PDF uncertainties, for a selection of processes in scenario $a_1$ (left panel of left figure; see also table~\ref{tab:xsec:a1a2}), $b_1$ (right panel of left figure; see also table~\ref{tab:xsec:b1b2}), and $c_1$ (right figure; see also table~\ref{tab:xsec:c1c2}). Predictions are normalised to the cross section obtained with the central {\tt CT18NLO} PDF set.}
    \label{fig:charmparametrisation}
\end{figure}

Due to the inclusion of $t$-channel and off-diagonal contributions, leptoquark pair production becomes more strongly dependent on the specific quark flavours triggered by the structure of the associated Yukawa couplings. Consequently, the way the charm density is accounted for in parton density fits may impact predictions for scenarios where leptoquarks interact with second-generation quarks. This is in particular relevant to understand the large differences observed between predictions obtained with the CT18/MSHT20 sets and those obtained with the NNPDF4.0 set. In order to further explore this feature, it must be noted that the NNPDF collaboration independently modeled an intrinsic charm contribution in their NNPDF4.0 baseline fit, and that it has alternatively released a set where the charm density purely arises perturbatively~\cite{Ball:2021leu}. In the present section, we discuss the effect of these different NNPDF parametrisations of the charm quark distribution on the production rates, together with the associated uncertainties. 

In figure~\ref{fig:charmparametrisation}, we display cross section predictions for a selection of processes from tables~\ref{tab:xsec:a1a2}--\ref{tab:xsec:c1c2}, for various PDF sets, and after including the corresponding PDF uncertainties. All results are normalised to predictions obtained with the central {\tt CT18NLO} set. In addition to the three PDF sets considered so far, we additionally evaluate the different production cross sections with the NNPDF4.0 set involving a perturbative treatment of the charm (\texttt{NNPD40\_nlo\_pch\_as\_01180}). Our results show that predictions obtained with the CT18, MSHT20 and NNPDF4.0 with perturbative charm sets all agree within their uncertainties, except for tensions arising in results for scenario $c_1$. In contrast, NNPDF4.0 predictions with intrinsic charm are significantly larger than those obtained with any of the other sets of parton densitites, the difference reaching a factor of 4--5 in the most extreme cases. The choice made for the charm parametrisation in the PDF fit has hence a large impact on our predictions.
%%%%%%%%%%%%%%%%%%%%%%%%%%%%%%%%%%
\section{Conclusions}
\label{ref:conclusions}
%%%%%%%%%%%%%%%%%%%%%%%%%%%%%%%%%%

In this work, we have studied diagonal and off-diagonal production modes of scalar leptoquark pairs at the LHC, taking into account NLO-QCD corrections and Yukawa-induced $t$-channel contributions. In particular, we have focused on generic simplified models involving either $R_2$ and $S_3$ leptoquarks or $R_2$ leptoquarks alone, as well as specific benchmark scenarios in $R_2$, $R_2$--$S_3$, and $S_1$--$S_3$ models addressing the flavour anomalies.

We find that in the cases studied here LO predictions are not reliable on their own. In order to obtain reliable theoretical predictions NLO corrections must be included. First, NLO corrections and the corresponding $K$-factors can be very big, both for diagonal and off-diagonal production. Their exact size depends of course on the specific process and the value of the Yukawa coupling, but they can reach from tens of percent up to a factor of $2$ for $m_{\rm LQ} = 1600~{\rm GeV}$ and a factor of about $4$ for $m_{\rm LQ} = 2400$ GeV. Secondly, LO predictions show a very strong dependence on the PDF sets used for the calculation. Adding NLO corrections dramatically improves the agreement between predictions obtained with different PDF sets. Furthermore, NLO corrections lead to  reduced  theoretical uncertainties both through the reduction of the scale dependence and the PDF error.

As can be expected, production rates for diagonal and off-diagonal modes grow with the value of the Yukawa coupling. While off-diagonal production is purely Yukawa coupling-induced (and therefore negligible at small values of the Yukawa coupling), diagonal leptoquark pair production involves $t$-channel contributions in addition to the pure QCD diagrams. The relevance of off-diagonal with respect to diagonal production including $t$-channel diagrams depends on the model, the process and the involved couplings. If a given off-diagonal process can proceed via two valence quarks in the initial state, it dominates at large values of the Yukawa coupling, with production rates much higher than for diagonal production (even with $t$-channel diagrams included). However, if there is no such PDF enhancement possible, at large values of the Yukawa coupling the rates for the off-diagonal production  are comparable or even smaller than those in the diagonal case. This clearly shows that in generic leptoquark models where large Yukawa couplings are relevant, in order to obtain reliable estimates of the pair-production rates at the LHC, one needs to consider -- with NLO precision -- not only the pure QCD cross sections but also the diagonal and off-diagonal $t$-channel contributions. Consequently, we urge both experimental and theoretical groups to use NLO predictions where these contributions are included instead of the LO ones.

We have also studied the relevance of the off-diagonal channels in a few scenarios addressing the flavour anomalies. We have found that for the benchmark points considered the off-diagonal production rates are in general small compared to the diagonal production. Again, the exact values depend on the model and the benchmark point but the effect ranges from negligible to ${\cal O}(10\%)$ of the total rate. In scenarios where leptoquarks couple to charm quarks, we observe that the different treatments applied to the charm quark distribution (perturbative or intrinsic) have a large impact on the theoretical predictions at NLO through the $t$-channel contributions, affecting in particular off-diagonal production.

\section*{Acknowledgements} 
The work of AJ is supported in part by a KIAS Individual Grant No.\ QP084401 via the Quantum Universe Center at Korea Institute for Advanced Study and by the Institute for Basic Science (IBS) under the project code, IBS-R018-D1. The work of BF has been partly supported by the French Agence Nationale de la Recherche (ANR) under grant ANR-21-CE31-0013 (project DMwithLLPatLHC). AK acknowledges the support and  hospitality of the CERN Theoretical Physics Department. The results of this study were obtained using the High Performance Computing facilities at the Center for Advanced Computation at KIAS. 
\appendix
\section{Leading-order amplitudes for off-diagonal leptoquark pair production}\label{app:amplitudes}

In this section, we consider the $R_2$--$S_3$ simplified model introduced in section~\ref{sec:benchmark_generic}, and we investigate the dependence of the various leptoquark off-diagonal pair production channels on the model's Yukawa couplings. We remind that all the entries of the coupling matrices are fixed to zero, except $y_{3,12}^{\rm LL}$, $y_{2,12}^{\rm LR}$, and $y_{2, 12}^{\rm RL}$. Moreover, we have imposed that all five leptoquark eigenstates are mass-degenerate, {\it i.e.}\ $m_{R_2} = m_{S_3} \equiv m_{\rm LQ}$. 

We consider generic off-diagonal leptoquark pair production processes,
\begin{eqnarray}
a(k_{1}, c_1)\quad b(k_{2}, c_2) \qquad\to\qquad {\rm LQ}_i(k_{3}, c_3) \quad {\rm LQ}_j(k_{4}, c_4),
\end{eqnarray}
where $a,b$ are the initial annihilating partons, $k_{m}$ with $m=1,2,3,4$ stand for the four-momenta of the different involved particles and $c_m$ are the corresponding colour indices. Moreover, in our notation, the generic leptoquark indices $i$ and $j$ only refer to the fact that the final-state leptoquarks are different. The $R_2$--$S_3$ model contains various off-diagonal channels that can be categorised into three categories according to the total electric charge of the final state.

\paragraph{Charge $\pm4/3$\! $-$}
\hspace{-.3cm}There are two partonic possibilities in the $R_2$--$S_3$ model to produce a pair of leptoquarks whose total electric charge is equal to $+4/3$,
\be 
   u u \to R_2^{(+2/3)}\ S_3^{(+2/3)} \qquad\text{and}\qquad uu\to R_2^{(+5/3)}\ S_3^{(-1/3)}\, .
\ee
These modes proceed through the annihilation of two up-type quarks via the $t$-channel exchange of either a charged lepton or a neutrino. The corresponding amplitudes are given by
\begin{equation}
\begin{split}
{\cal M}_{R_2^{(+2/3)} S_3^{(+2/3)}} &= - \sqrt{2} y_{2,12}^{\rm RL}\, y_{3,12}^{\rm LL} \bigg[(\bar{v}_1 P_L \slashed{k}_3 u_2) \frac{1}{\hat{u}} \delta_{c_1 c_4} \delta_{c_2 c_3} - (\bar{v}_1 P_R \slashed{k}_3 u_2) \frac{1}{\hat{t}} \delta_{c_1 c_3} \delta_{c_2 c_4} \bigg], \\
{\cal M}_{R_2^{(+5/3)} S_3^{(-1/3)}} &= y_{2,12}^{\rm LR}\, y_{3,12}^{\rm LL} \bigg[m_\mu \frac{1}{\hat{u}} (\bar{v}_1 P_L u_2) \delta_{c_1 c_4} \delta_{c_2 c_3} - (\bar{v}_1 P_L \slashed{k}_3 u_2) \frac{1}{\hat{u}} \delta_{c_1 c_4} \delta_{c_2 c_3} \\
&\qquad\qquad\quad~~ + m_\mu (\bar{v}_1 P_L u_2) \frac{1}{\hat{t}} \delta_{c_1 c_3} \delta_{c_2 c_4} +  \bar{v}_2 P_R \slashed{k}_3 u_1 \frac{1}{\hat{t}} \delta_{c_1 c_3} \delta_{c_2 c_4}\bigg]\,.
\end{split}
\label{eq:amplitudes:1}
\end{equation}
In these expressions, $\hat t$ and $\hat u$ denote usual Mandelstam variables, $\bar v_m$ and $u_m$ stand for four-component spinors associated with an initial particle of momentum $k_m$, $P_L$ and $P_R$ are left-handed and right-handed chirality projectors, and $m_\mu$ refers to the muon mass. 

After averaging over the initial degrees of freedom and squaring the amplitudes, we obtain the following relation between the associated cross sections
\begin{eqnarray}
\frac{\sigma\Big[pp\to R_2^{(+5/3)} S_3^{(-1/3)}\Big]}{\sigma\Big[pp\to R_2^{(+2/3)} S_3^{(+2/3)}\Big]} \approx \frac{1}{2} \bigg(\frac{y_{2,12}^{\rm LR}}{y_{2,12}^{\rm RL}}\bigg)^2\,,
\end{eqnarray}
which can also be derived at the level of the amplitudes~\eqref{eq:amplitudes:1} in the limit $m_\mu \to 0$. The complex conjugate processes lead to relatively much smaller cross sections when the leptoquarks couple to up quarks, whereas they are comparable when initial quarks are charm quarks. These effects are PDF-driven.

\paragraph{Charge $\pm1$\! $-$}
\hspace{-.3cm}Three partonic processes lead to the production of a pair of leptoquarks whose total electric charge is equal to $+1$,
\be
  \bar{d} u \to  R_2^{(-2/3)} R_2^{(+5/3)},\qquad
  \bar{d} u \to S_3^{(+1/3)} S_3^{(+2/3)}
  \qquad\text{and}\qquad
  \bar{d} u \to S_3^{(+4/3)} S_3^{(-1/3)}\,.
\ee
At leading order, this proceeds through the annihilation of a down-type antiquark and an up-type quark, and the corresponding amplitudes read
\begin{eqnarray}
{\cal M}_{R_2^{(-2/3)} R_2^{(+5/3)}} &=& 
   \Big(y_{2,12}^{\rm RL}\Big)^2\ \Big[m_\mu \bar{v}_1 P_R u_2 + \bar{v}_1 P_R \slashed{k}_3 u_2 \Big]\  \delta_{c_1 c_3} \delta_{c_2 c_4} \frac{1}{\hat{t}}\,, \nonumber \\
{\cal M}_{S_3^{(+1/3)} S_3^{(+2/3)}} &=& 
  \sqrt{2} \Big(y_{3,12}^{\rm LL}\Big)^2\ \Big[\bar{v}_1 P_R \slashed{k}_3 u_2 \Big]\ \delta_{c_1 c_3} \delta_{c_2 c_4} \frac{1}{\hat{t}}, \\
{\cal M}_{S_3^{(+4/3)} S_3^{(-1/3)}}  &=& \sqrt{2} \Big(y_{3,12}^{\rm LL}\Big)^2\ \Big[\bar{v}_1 P_R \slashed{k}_3 u_2\Big]\  \delta_{c_1 c_3} \delta_{c_2 c_4} \frac{1}{\hat{t}}. \nonumber
\label{eq:amplitudes:2}
\end{eqnarray}
From these expressions, we derive  the associated cross sections and find that they satisfy
\begin{eqnarray}
\frac{\sigma\Big[pp\to R_2^{(-2/3)} R_2^{(+5/3)}\Big]}{\sigma\Big[pp\to S_3^{(+1/3)} S_3^{(+2/3)}\Big]} = \frac{\sigma\Big[pp\to R_2^{(-2/3} R_2^{(+5/3)}\Big]}{\sigma\Big[pp\to S_3^{(+4/3)} S_3^{(-1/3)}\Big]} \approx \frac{1}{2} \bigg(\frac{y_{2,12}^{\rm RL}}{y_{3,12}^{\rm LL}}\bigg)^4\,.
\end{eqnarray}

\paragraph{Charge $\pm1/3$\! $-$}
\hspace{-.3cm}There are two partonic processes leading to the production of a pair of leptoquarks whose total electric charge is equal to $+1/3$,
\be
  u d \to R_2^{(+2/3)} S_3^{(-1/3)}\qquad\text{and}\qquad
  u d \to R_2^{(+5/3)} S_3^{(-4/3)}\,.
\ee
They occur through the annihilation of an up-type and a down-type quark, and in the limit of vanishing muon mass, the corresponding Feynman amplitudes read
\be\bsp
{\cal M}_{R_2^{(+2/3)} S_3^{(-1/3)}} =&\ - \Big(y_{2,12}^{\rm RL} y_{3,12}^{\rm LL}\Big)\ \Big[\bar{v}_1 P_L \slashed{k}_3 u_2\Big]\  \delta_{c_1 c_3} \delta_{c_2 c_4} \frac{1}{\hat{t}}\,, \\
{\cal M}_{R_2^{(+5/3)} S_3^{(-4/3)}} =&\ \sqrt{2} \Big(y_{2,12}^{\rm RL} y_{3,12}^{\rm LL}\Big)\ \Big[\bar{v}_1 P_L \slashed{k}_3 u_2\Big]\ \delta_{c_1 c_3} \delta_{c_2 c_4} \frac{1}{\hat{t}}\,.
\esp\label{eq:amplitudes:3}\ee
Using the fact these two processes have the same partonic luminosities, one derive a ratio of cross sections that fulfils 
\begin{eqnarray}
\frac{\sigma\Big[pp\to R_2^{(+2/3)} S_3^{(-1/3)}\Big]}{\sigma\Big[pp\to R_2^{(+5/3)} S_3^{(-4/3)}\Big]} = \frac{1}{2}\,.\label{r2s3scaling}
\end{eqnarray}

\begin{figure}[!tbp]
    \centering
    \includegraphics[width=0.85\linewidth]{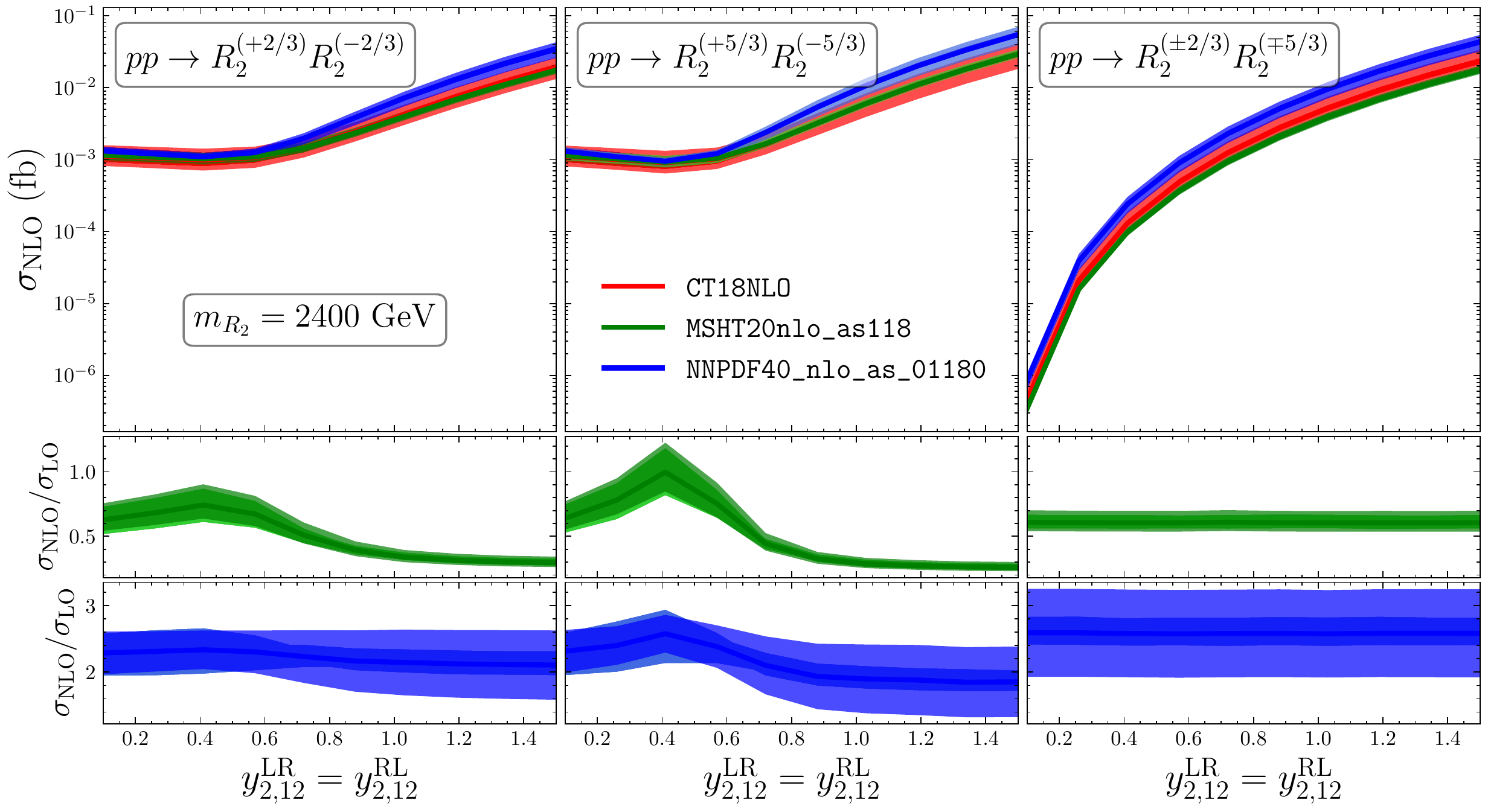}
    \caption{Same as in figure \ref{fig:xsec:R2R2:CH1:y12:MLQ:1600} but for $m_{R_2} = 2400~{\rm GeV}$.}
    \label{fig:xsec:R2R2:CH1:y12:MLQ:2400}
\end{figure}

\begin{figure}[!h]
\centering
\includegraphics[width=0.85\linewidth]{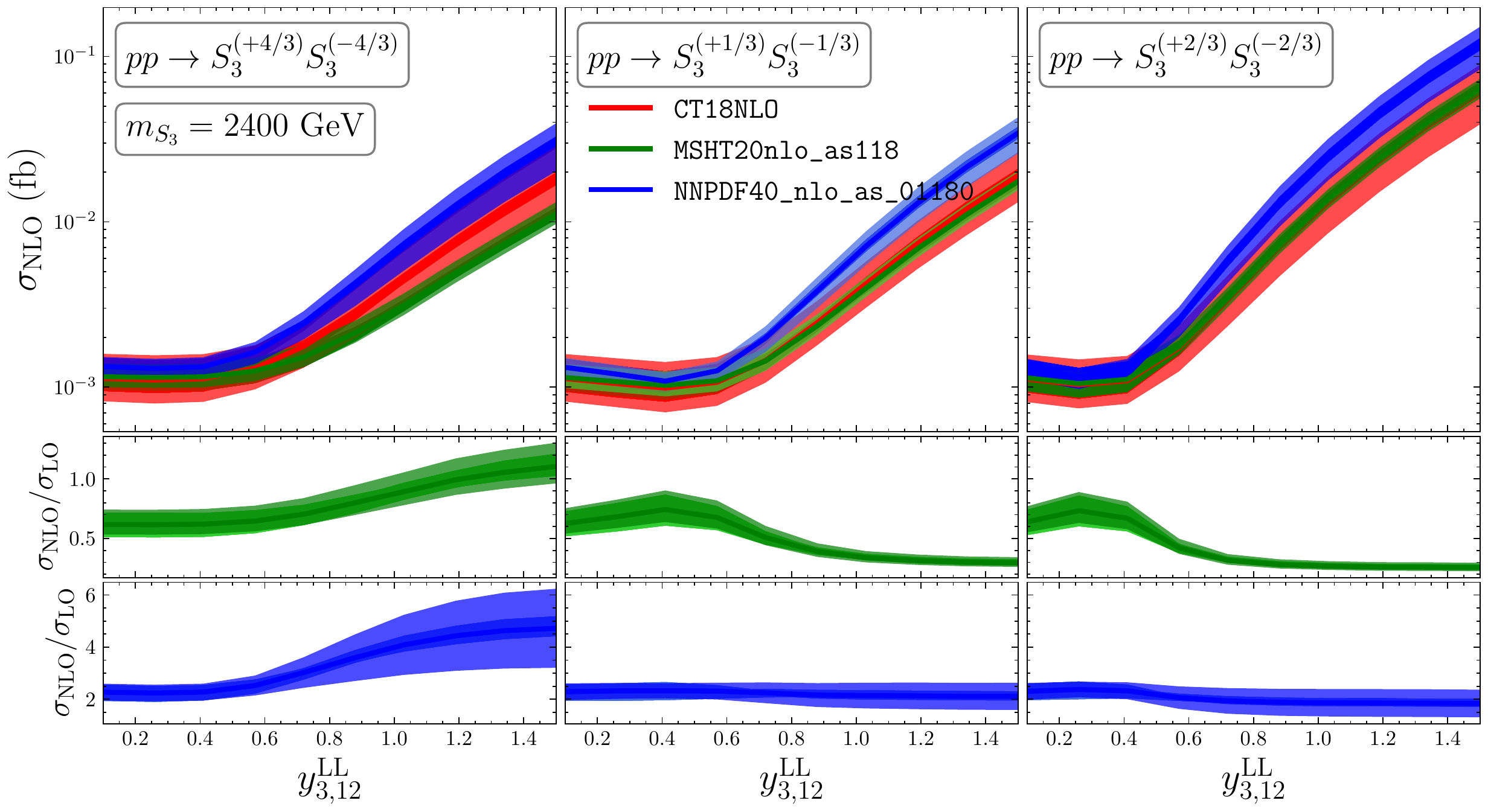}\\
\includegraphics[width=0.6\linewidth]{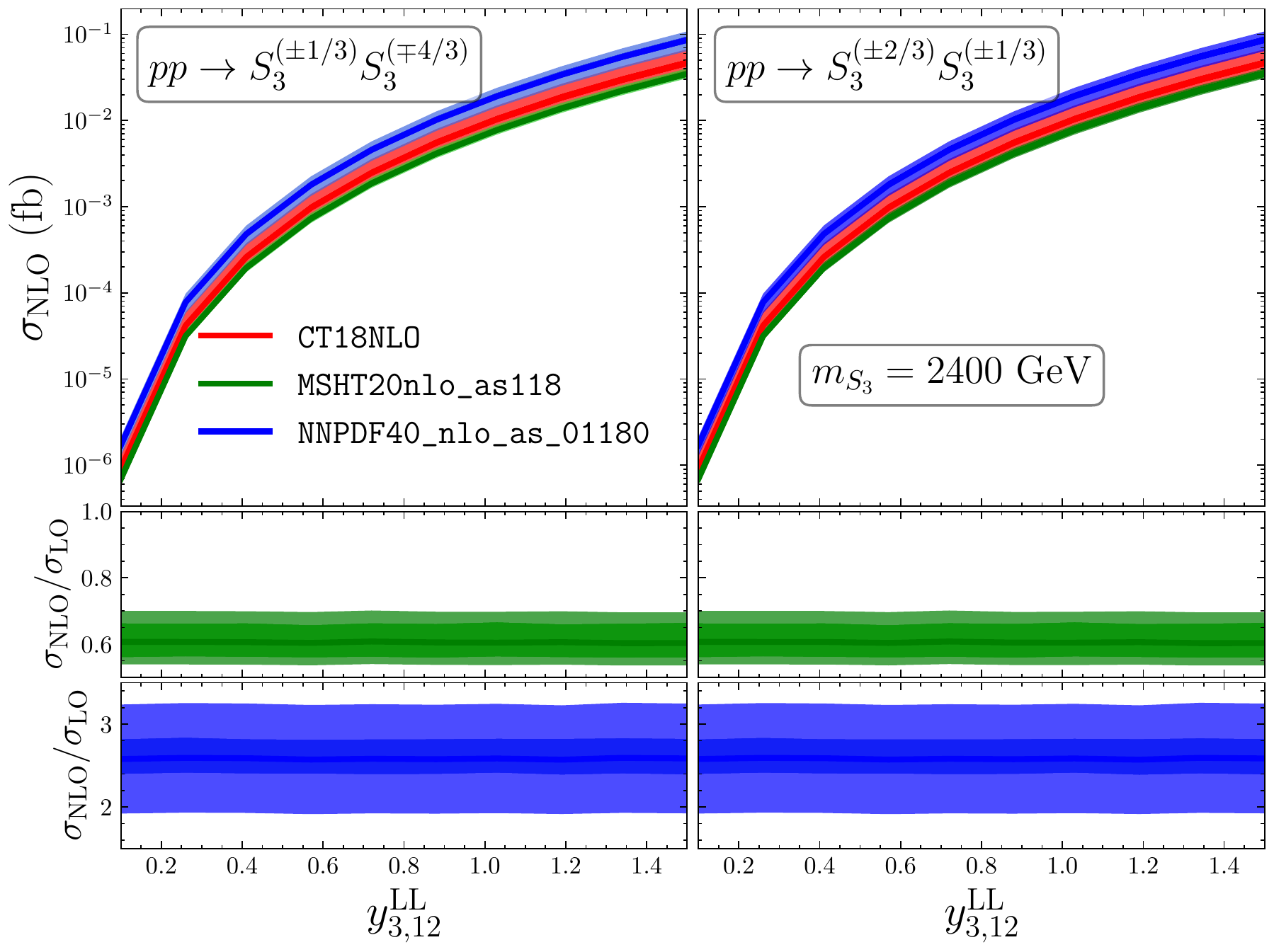}
\caption{Same as in figure \ref{fig:xsec:S3S3:CH2:y12:MLQ:1600} but for $m_{R_2} = m_{S_3} = 2400~{\rm GeV}$.}
\label{fig:xsec:S3S3:CH1:y12:MLQ:2400}
\end{figure}
%%%%%%%%%%%%%%%%%%%%%%%%%%%%%%%%%%%%%%%%%%%%%%%%
%%%%%%%%%%%%%%%%%%%%%%%%%%%%%%%%%%%%%%%%%%%%%%%%

\section{Additional results for $m_{\rm LQ} = 2400~{\rm GeV}$}
\label{app:highermass}

In this section, we show predictions for leptoquark production in the $R_2$--$S_3$ model for $m_{R_2} = m_{S_3} = 2400~{\rm GeV}$. Figure~\ref{fig:xsec:R2R2:CH1:y12:MLQ:2400} shows total cross sections at NLO (and the corresponding $K$-factors) for the processes $pp\to R_2^{(+2/3)} R_2^{(-2/3)}$, $pp\to R_2^{(+5/3)} R_2^{(-5/3)}$ and $pp\to R_2^{(\pm5/3)} R_2^{(\mp2/3)}$, and investigates how they depend on $y = y_{2,12}^{\rm LR} = y_{2,12}^{\rm RL}$. Our findings are similar as for $m_{R_2} = 1600$ GeV (see figure \ref{fig:xsec:R2S3:CH1:y12:MLQ:1600}), although predictions obtained with NNPDF4.0 suffer from much larger errors. When increasing $y$ from $0.8$ to $1.5$, we observe the resulting diagonal production mode cross sections to increase by almost a factor of $30$, reaching about $\mathcal{O}(50)~{\rm ab}$. Additionally, the off-diagonal production rate also increases by almost five orders of magnitudes, being similar to the diagonal rates for Yukawa coupling values of $y = 1.5$. Finally, whereas LO predictions depend quite a lot on the chosen PDF set (as shown with the $K$-factors), this dependence is reduced at NLO. Similar trends can be observed for processes involving a diagonal pair of $S_3$ leptoquark eigenstates in figure~\ref{fig:xsec:S3S3:CH1:y12:MLQ:2400}.

\begin{figure}[!t]
    \centering
    \includegraphics[width=0.85\linewidth]{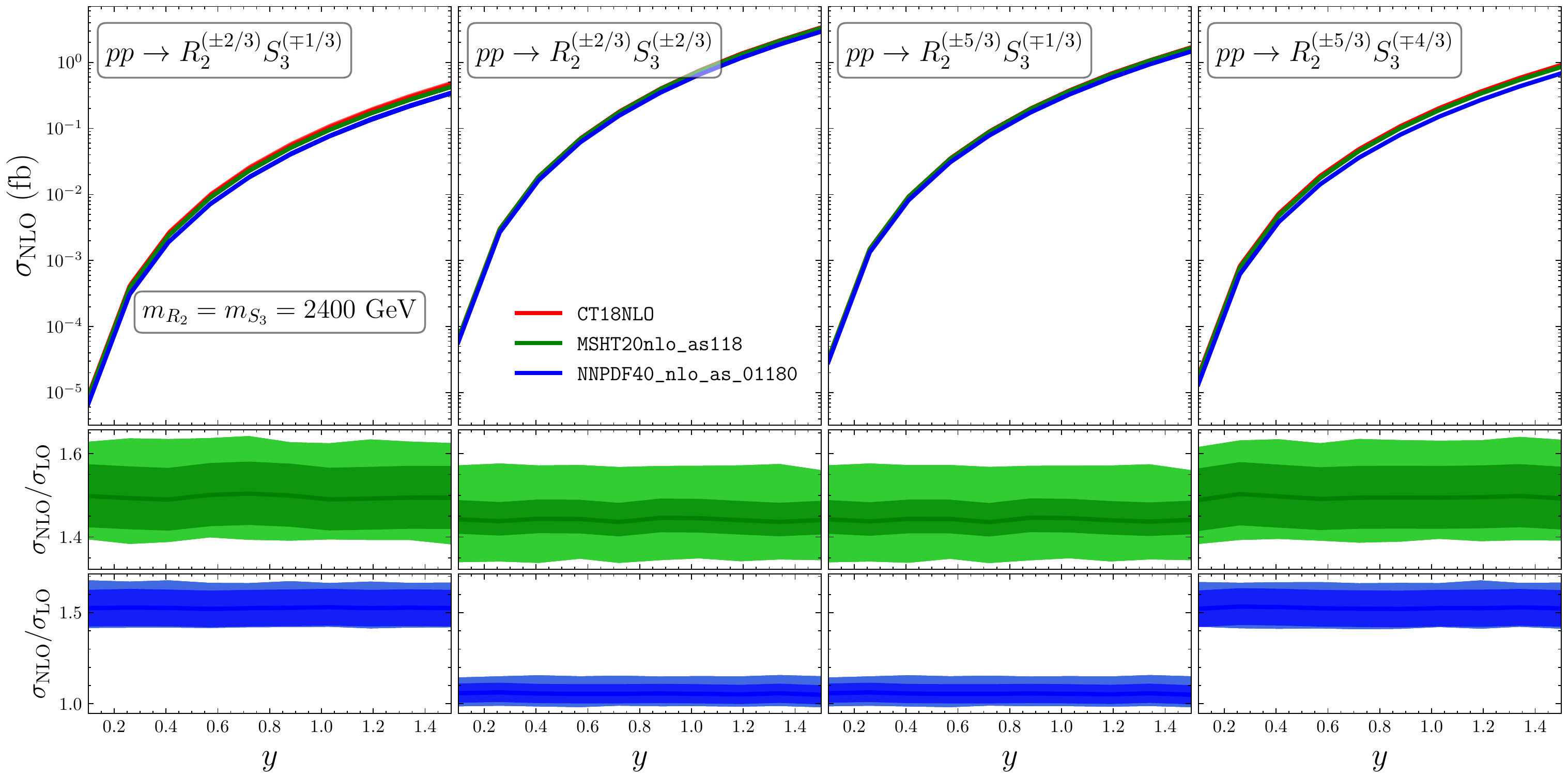}
    \caption{Same as in figure \ref{fig:xsec:R2S3:CH1:y12:MLQ:1600} but for $m_{R_2} = m_{S_3} = 2400~{\rm GeV}$.}
    \label{fig:xsec:R2S3:CH1:y12:MLQ:2400}
\end{figure}

Results for the mixed $R_2$--$S_3$ channels ($pp\to R_2^{(\pm2/3)}S_3^{(\mp1/3)}$, $pp\to R_2^{(\pm2/3)}S_3^{(\pm2/3)}$, $pp\to R_2^{(\pm5/3)}S_3^{(\mp1/3)}$, and $R_2^{(\pm5/3)}S_3^{(\mp4/3)}$) are shown in figure~\ref{fig:xsec:R2S3:CH1:y12:MLQ:2400}. As for all the off-diagonal channels, $K$-factors are flat and independent of the value of the $y$ coupling. On the other hand, the size of these $K$-factors are process-dependent for the NNPDF4.0 PDF set, whilst they are almost independent of the process for the MSHT20 PDF set. This can be again traced back to the poor PDF fits at LO, an effect that is fully compensated with NLO predictions. Moreover, rates for the mixed channels increase dramatically for large $y$ values, reaching values of $1$--$3$ fb that are relevant for the LHC for the $pp\to R_2^{(\pm5/3)} S_3^{(\mp4/3)}$, $pp\to R_2^{(\pm5/3)} S_3^{(\mp1/3)}$, and $pp\to R_2^{(\pm2/3)} S_3^{(\pm2/3)}$ processes.

\begin{figure}
    \centering
    \includegraphics[width=0.49\linewidth]{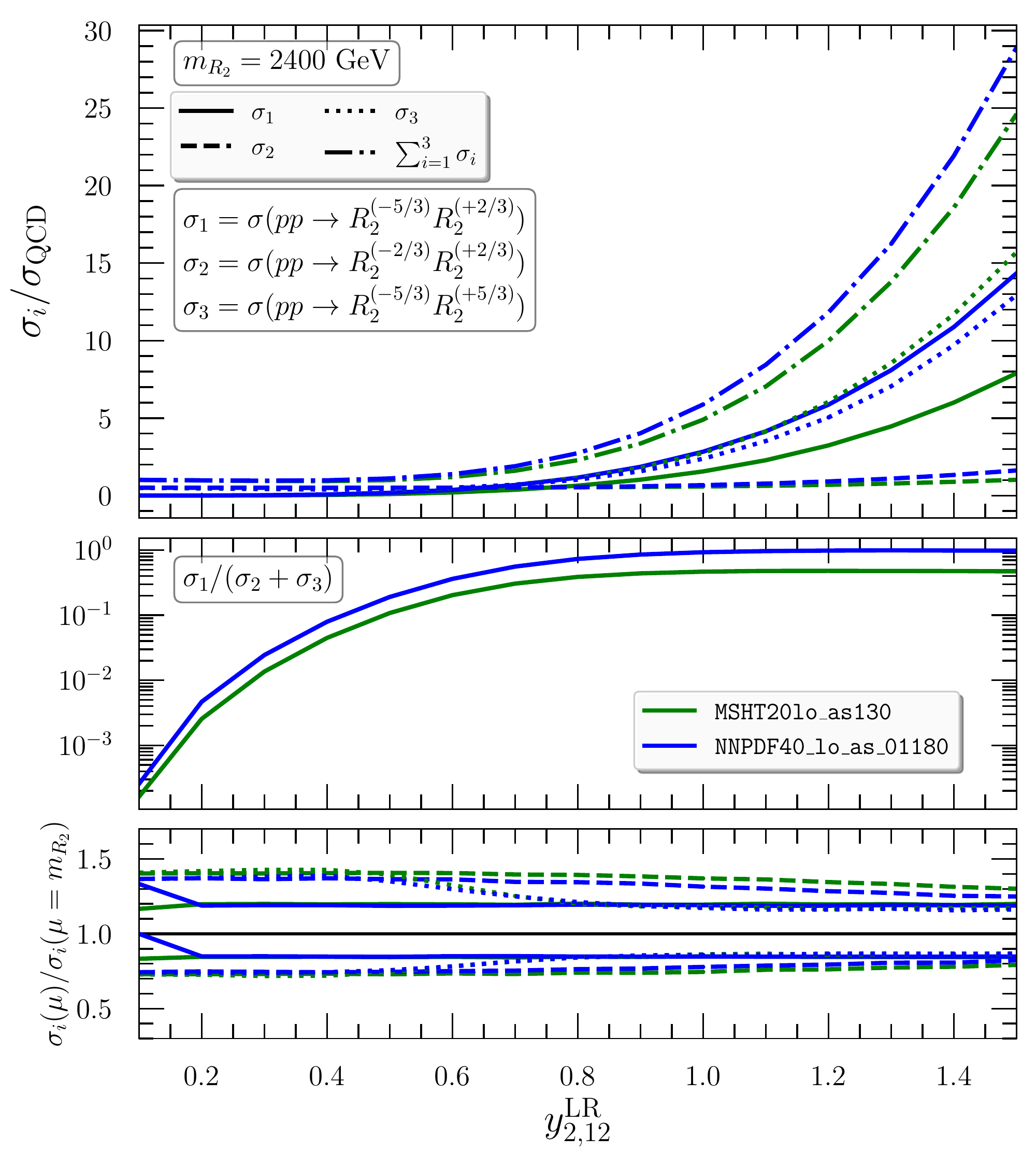}
    \hfill 
    \includegraphics[width=0.49\linewidth]{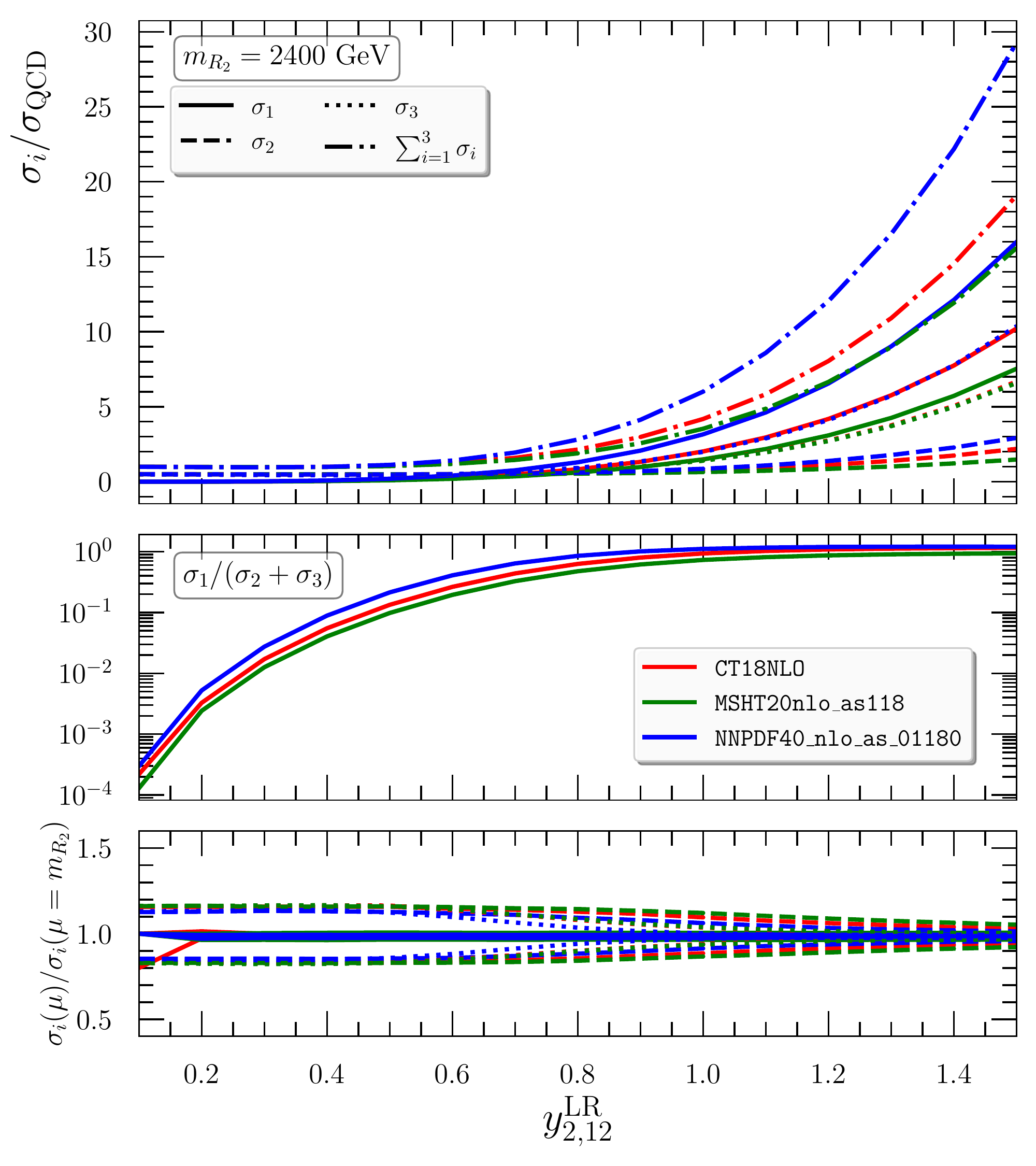}
    \caption{Same as in figure \ref{fig:ratio:R2:y12:MLQ:1600} but for $m_{R_2} = 2400$ GeV.}
    \label{fig:ratio:R2:y12:MLQ:2400}
    \vspace{.4cm}
    \centering
    \includegraphics[width=0.49\linewidth]{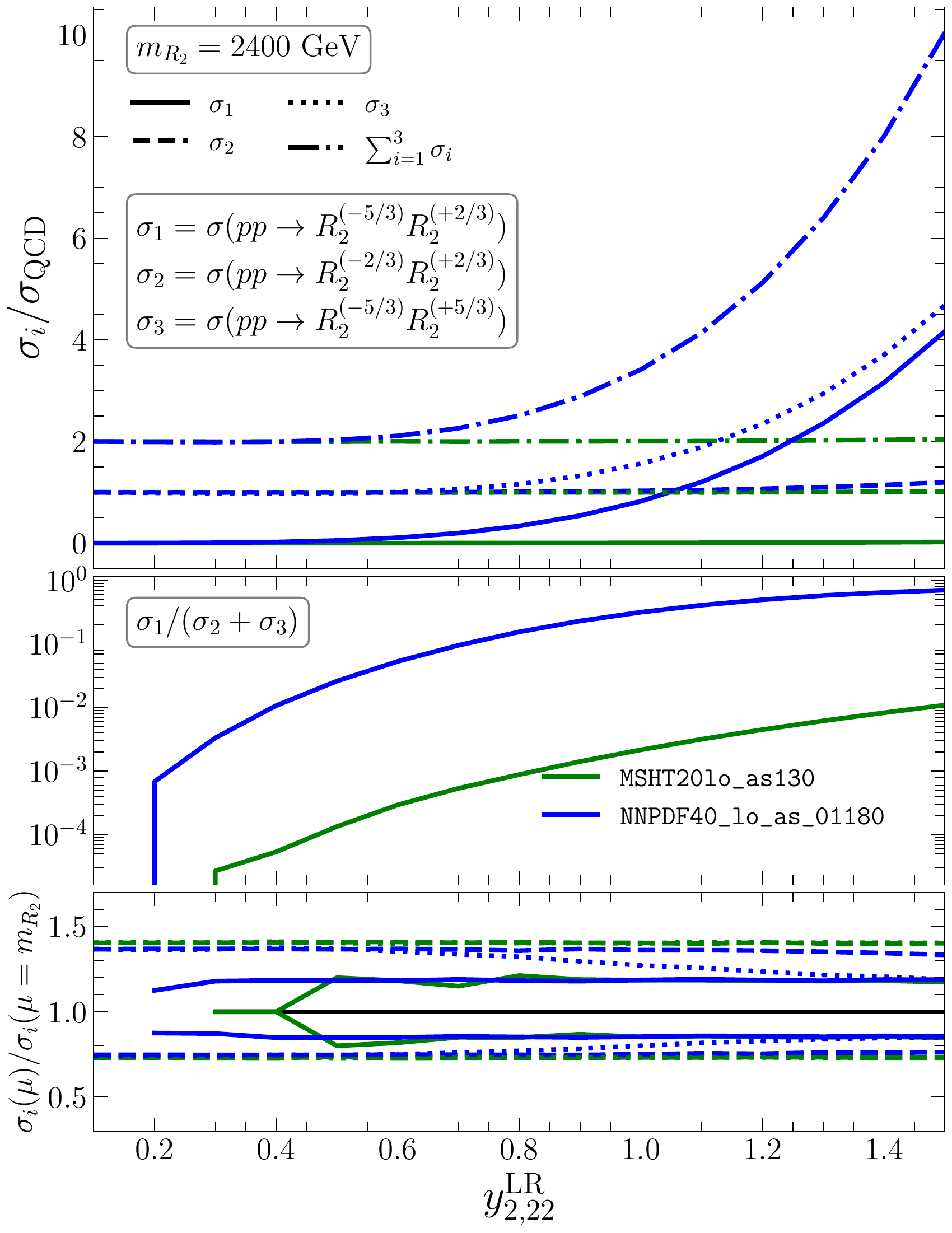}
    \hfill 
    \includegraphics[width=0.49\linewidth]{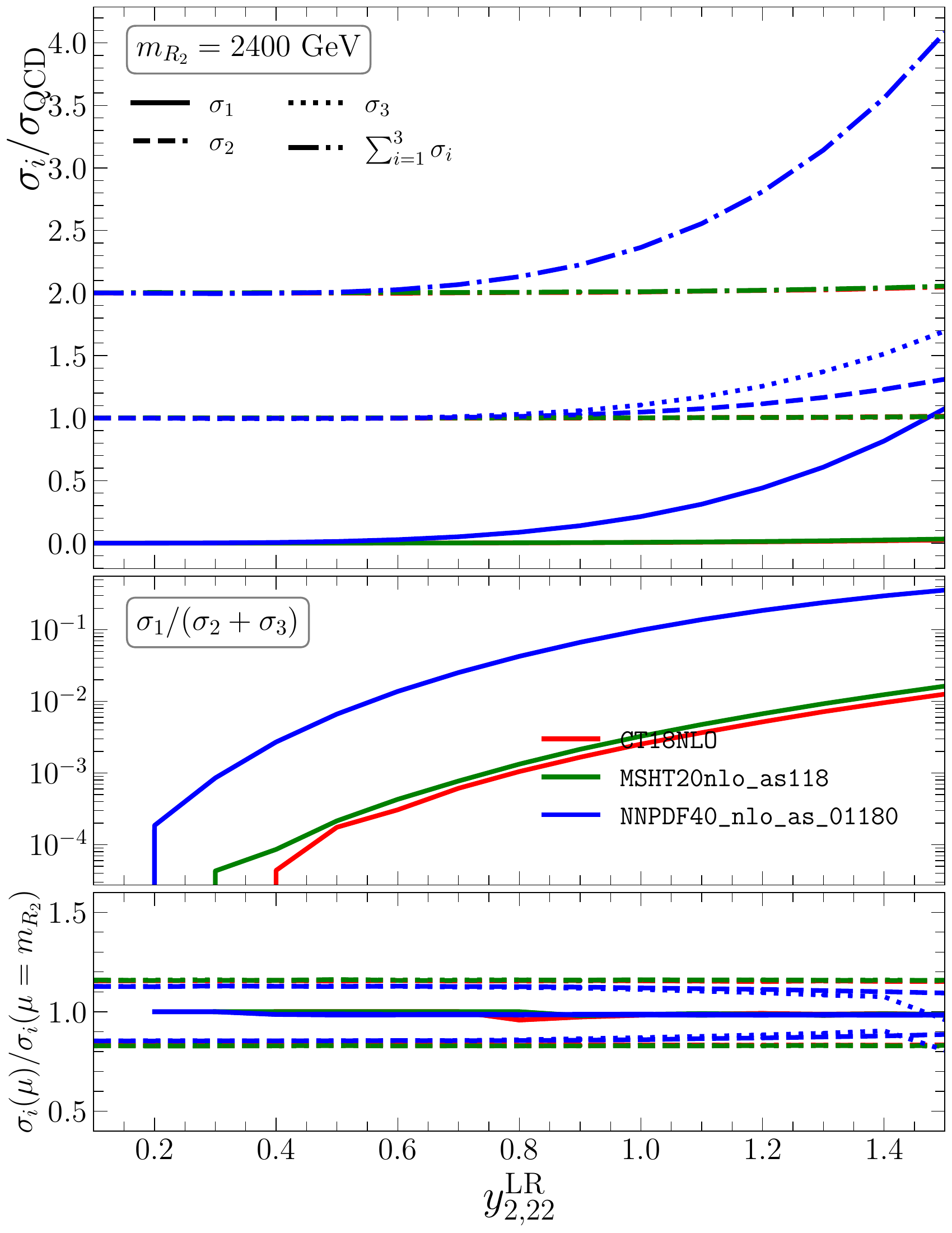}
    \caption{Same as in figure \ref{fig:ratio:R2:y22:MLQ:1600} but for $m_{R_2} = 2400$ GeV.}
    \label{fig:ratio:R2:y22:MLQ:2400}
\end{figure}

\begin{figure}
    \centering
    \includegraphics[width=0.49\linewidth,height=12cm]{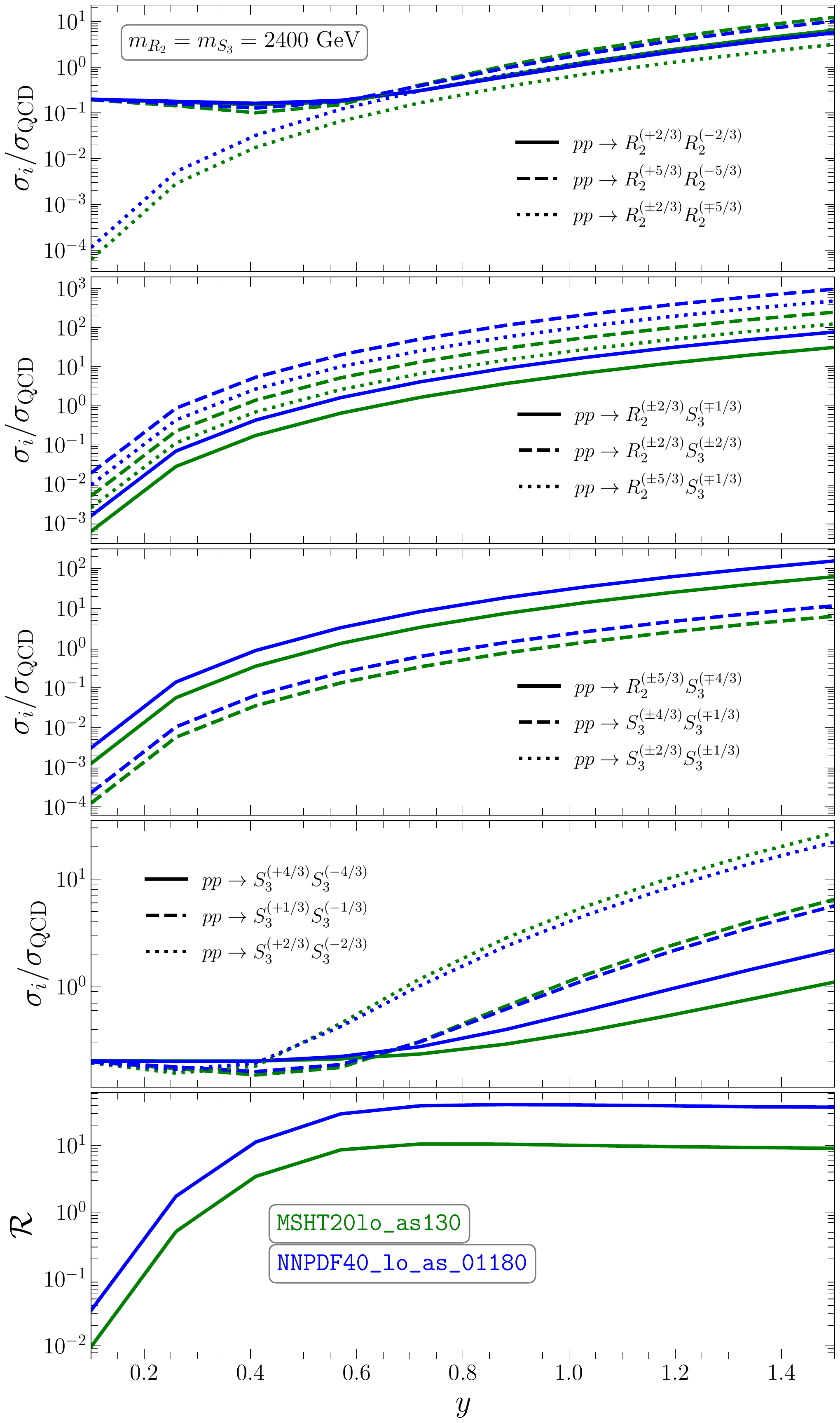}
    \hfill
    \includegraphics[width=0.49\linewidth,height=12cm]{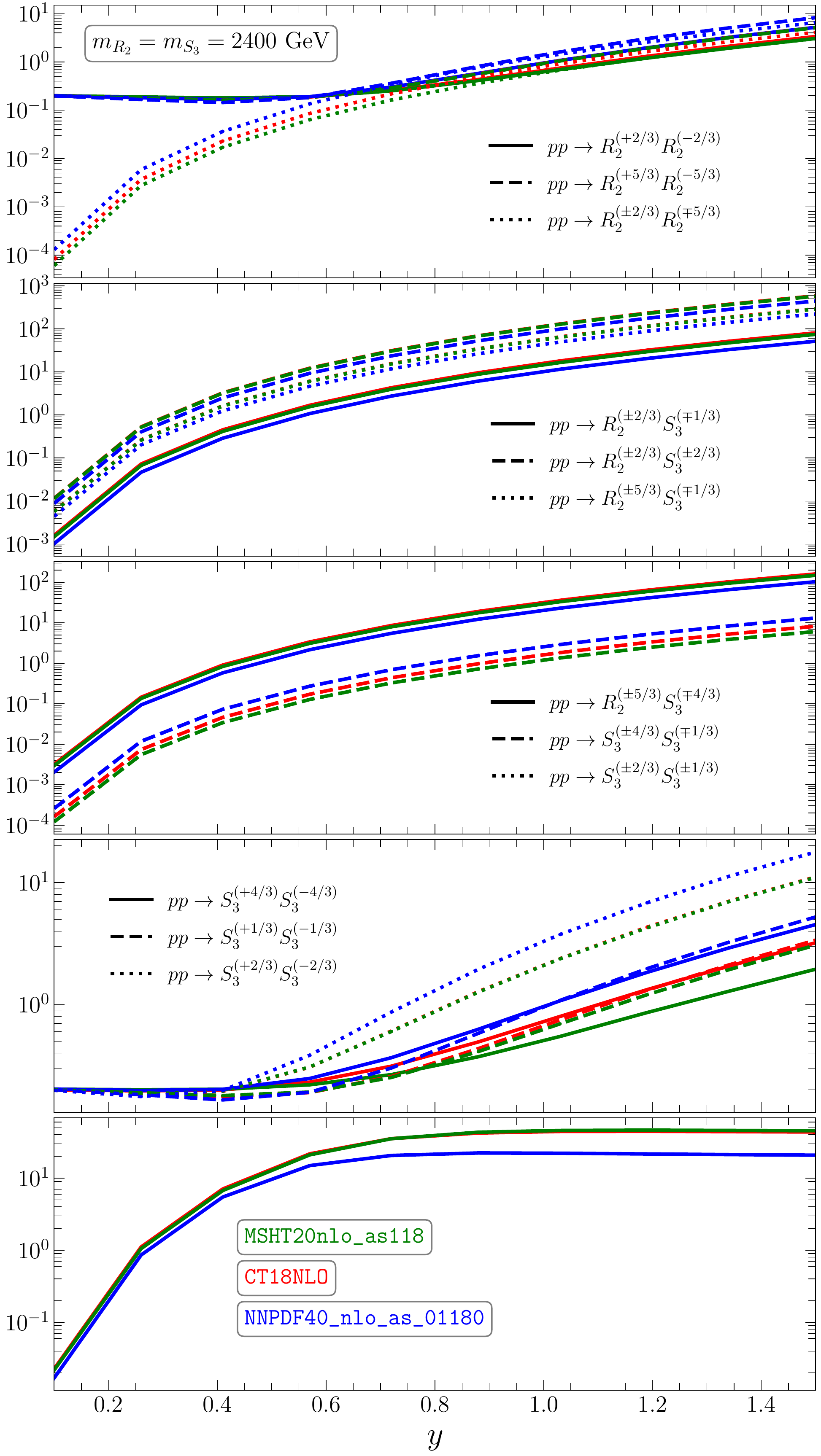}
    \caption{Same as in figure \ref{fig:ratios:R2S3:mLQ:1600} but for $m_{R_2} = m_{S_3} = 2400~{\rm GeV}$.}
    \label{fig:ratios:R2S3:mLQ:2400}
\end{figure}

Results illustrating the importance of the off-diagonal and $t$-channel contributions over the pure 
QCD contribution for $m_{\rm LQ} = 2400$ GeV are reported in figures~\ref{fig:ratio:R2:y12:MLQ:2400}--\ref{fig:ratios:R2S3:mLQ:2400}. These figures show similar findings as those described in section~\ref{sec:tchannel:MLQ:1600}.

\bibliographystyle{JHEP}
\bibliography{bibliography.bib}

%%%%%%%%%%%%%%%%%%%%%%%%%%%%

\end{document}